\documentclass[12pt,a4paper]{article}
\usepackage{graphicx}
\usepackage{amsmath}
\usepackage{amssymb}
\usepackage{bbold}
\usepackage{type1cm}
\usepackage{rotating}
\usepackage{tabularx}
\usepackage{pdflscape}

\usepackage{dsfont}
\usepackage{overpic}
\usepackage{cite}
\usepackage{slashed}
\usepackage{bbm}
\usepackage{subfigure}
\usepackage{graphicx,textcomp}
\allowdisplaybreaks

\author{Peter C. Bruns}
\title{A formalism for the study of $K^{+}\pi\,\Sigma$ photoproduction in the $\Lambda^{\ast}(1405)$ region}
\date{\vspace{-5ex}}

\begin{document}
\maketitle

\begin{center}
Nuclear Physics Institute of the Czech Academy of Sciences, 25068 \v{R}e\v{z}, Czech Republic
\end{center}

\quad \\
\begin{center}
  {\bf Abstract}\\
\end{center}
We present a formalism for the low-energy analysis of the $\gamma p\rightarrow K^{+}\pi\Sigma$ photoproduction reaction in the $\Lambda^{\ast}(1405)$ resonance region. In particular, the constraints arising from unitarity, gauge invariance and chiral perturbation theory are discussed in some detail.\\



\maketitle

\section{Introduction}

Just below the antikaon-nucleon threshold, one finds a strong enhancement of the strangeness $S=-1$ meson-baryon interaction, which is ascribed to the $\Lambda^{\ast}(1405)$ $I(J^{P})=0(\frac{1}{2}^{-})$ resonance. Its existence was conjectured by Dalitz and Tuan in 1959 \cite{Dalitz:1959dn,Dalitz:1959dq,Dalitz:1960du} and soon after confirmed in the analysis of $\pi\Sigma$ mass spectra observed in bubble chamber experiments \cite{Alston:1961zzd,Bastien:1961zze}. We refer to recent reviews on this resonance \cite{Hyodo:2011ur,Mai:2020ltx} for more references, an up-to-date overview of theoretical and experimental methods, and a discussion of the impact of the $\Lambda^{\ast}(1405)$ on various areas of hadron physics and astrophysics. \\
Applying modern approaches to the coupled-channel meson-baryon scattering problem, where interaction potentials derived from chiral Lagrangians are iterated to an infinite order in the loop expansion, it was found \cite{Oller:2000fj,GarciaRecio:2002td,Jido:2003cb} that there are {\em two\,} poles on the complex-energy Riemann surface of the scattering amplitude that correspond to the $\Lambda^{\ast}(1405)$ (as discussed recently in \cite{Meissner:2020khl}, this might be a more common phenomenon in hadron physics than one could have thought). In \cite{Jido:2003cb}, it was furthermore demonstrated that one of the poles is related to a flavor-octet state, while the other one is related to a flavor-singlet state. However, while the position of the octet-related pole, which is located closer to the $\bar{K}N$ threshold, seems to be well fixed by the experimental data and the theoretical approach, there is apparently a considerable model dependence involved in the determination of the broader, presumably singlet-related pole (compare e.g. Fig.~7 in \cite{Mai:2020ltx}, and the discussions in \cite{Cieply:2016jby,Dong:2016auh,Revai:2017isg,Myint:2018ypc,Bruns:2019bwg,Anisovich:2020lec}).\\
A possible way to constrain this model dependence is provided by the measurement of the $\gamma p\rightarrow K^{+}\pi\Sigma$ photoproduction cross sections by the CLAS collaboration. In \cite{Moriya:2013eb}, the $\pi\Sigma$ line shapes in the $\Lambda^{\ast}(1405)$ resonance region were determined, and in \cite{Moriya:2013hwg}, also the $K^{+}$ angular distributions were measured (see also \cite{Ahn:2003mv,Niiyama:2008rt,Lu:2013nza}). The pole structure is expected to influence the $\pi\Sigma$ line shapes notably, so that it should be possible to rule out classes of models which are not consistent with the CLAS data. An interesting approach was followed in \cite{Roca:2013av,Roca:2013cca,Mai:2014xna}, where the elementary photoproduction amplitude was parameterized by a set of constants $C^{c}(\sqrt{s})$ (for each reaction channel $c$, and each bin of the c.m. energy $\sqrt{s}$). Multiplying this by a meson-baryon loop function and a unitarized coupled-channel meson-baryon scattering amplitude to describe the final-state interaction, one obtains a simple model parameterization for the full production amplitude.  Comparing the quality of fits to the CLAS data with this class of models then allows to test the employed scattering amplitudes. The drawbacks of such an approach are, of course, that it is hard to judge whether the obtained fit results for the $C^{c}(\sqrt{s})$ make good physical sense, and that one does not arrive at a more detailed understanding of the full process. To achieve the latter, microscopic models for the elementary photoproduction amplitude, based on effective Lagrangians, have been developed \cite{Nacher:1998mi,Nakamura:2013boa} (see also \cite{Lutz:2004sg,Wang:2016dtb,Nam:2017yeg}). It is our aim to find a compromise between these two classes of approaches, by allowing some flexibility of the elementary photoproduction amplitude, but on the other hand implementing rigorous low-energy constraints derived from chiral perturbation theory (ChPT), the chiral effective field theory of QCD. Since this project requires some amount of formalism as well as detailed explanation of fit procedures and data analysis, we decided to split the publication of this ongoing work in several parts. The contribution at hand contains a detailed and explicit presentation of the formalism to be employed, while the forthcoming studies based on that formalism will be a bit less technical, but more concerned with the experimental data to be described. \\
This article is organized as follows. In Sec.~\ref{sec:generalities}, we start with some generalities like kinematics, the decomposition of the photoproduction amplitude into invariant Lorentz structures, and the formula for the cross section. In Sec.~\ref{sec:LETs}, we present the low-energy constraints as derived from ChPT. While the tree graph topologies we include have substantial overlap with those in \cite{Nakamura:2013boa}, we avoid non-relativistic approximations, and give the full results explicitly in App.~\ref{app:treegraphs}. The ensuing Sec.~\ref{sec:twostep} contains a short but (hopefully) instructive digression dealing with explicit resonance degrees of freedom and Breit-Wigner approximations. It is mainly meant as a preparation for Sec.~\ref{sec:cc_formalism}, where we explain our approach to the coupled-channel unitarization of the photoproduction amplitude, which is somewhat different from the method used in \cite{Nacher:1998mi,Nakamura:2013boa}, but more akin to the one of \cite{Kaiser:1996js}. Again, this can be related to our manifestly covariant treatment of the amplitudes. Finally, Sec.~\ref{sec:summary} gives a short summary and outlook.

\section{Kinematics, cross section, and general formalism}
\label{sec:generalities}

We consider the reaction $\gamma(k)p(p_{N})\,\rightarrow\,K(q_{K})\pi(q_{\pi})\Sigma(p_{\Sigma})$\,, where the symbol in brackets denotes the four-momentum of the indicated particle. Throughout, we shall neglect isospin-symmetry breaking corrections, so e.g. proton and neutron have a common nucleon mass $m_{N}$, and all $\Sigma$ hyperons a common mass $m_{\Sigma}$.  For a process with five external particles, we can form $\left(\begin{array}{c} 5-1 \\ 2 \end{array}\right)-1 = 5$ independent Mandelstam variables, which we choose here as
\begin{eqnarray}
  s &=& (p_{N}+k)^2 = (q_{K}+q_{\pi}+p_{\Sigma})^2\,, \hspace{-0.2cm}\qquad M_{\pi\Sigma}^2=(q_{\pi}+p_{\Sigma})^2\,,\nonumber \\
  t_{\Sigma} &=& (p_{\Sigma}-p_{N})^2\,, \quad u_{\Sigma} = (p_{\Sigma}-k)^2\,, \quad t_{K} = (q_{K}-k)^2\,.\label{eq:mandelstams} 
\end{eqnarray}
They are chosen in such a way that in the ``soft pion limit'' $q_{\pi}\rightarrow 0$, $s,t_{\Sigma}\rightarrow t_{K}=:t$ and $u_{\Sigma}=:u$ become the usual variables $s,t,u$ of single-kaon photoproduction. We note that the squared invariant masses for the other two-particle channels in the final state can be expressed as
\begin{eqnarray}
  M_{\pi K}^2 &:=& (q_{\pi}+q_{K})^2 = s+t_{\Sigma}+u_{\Sigma}-m_{\Sigma}^2-m_{N}^2-k^2\,,\\
  M_{K\Sigma}^2 &:=& (q_{K}+p_{\Sigma})^2 = m_{\Sigma}^2+M_{K}^2+m_{N}^2+k^2-t_{\Sigma}-u_{\Sigma}-(M_{\pi\Sigma}^2-m_{\Sigma}^2-M_{\pi}^2)\,.
\end{eqnarray}
We also note that we only deal with photoproduction in this work, so $k^2\equiv k_{\mu}k^{\mu}=0$. In the overall c.m. frame, where $\vec{p}_{N}+\vec{k}=\vec{0}$, we find, employing energy-momentum conservation,
\begin{equation}\label{eq:qK}
q_{K}^{0} \equiv E_{K} = \frac{s+M_{K}^2-M_{\pi\Sigma}^2}{2\sqrt{s}} = \sqrt{|\vec{q}_{K}|^2+M_{K}^2}\,,\quad |\vec{q}_{K}|=\frac{\sqrt{\lambda(s,M_{\pi\Sigma}^2,M_{K}^2)}}{2\sqrt{s}}\,,
\end{equation}
where $\lambda(x,y,z):=x^2+y^2+z^2-2xy-2xz-2yz$\,, while in the $\pi\Sigma$ c.m. frame (which we shall indicate by a $\ast$ superscript), the related quantities are given by
\begin{equation}\label{eq:qKast}
E_{K}^{\,\ast} = \frac{s-M_{K}^2-M_{\pi\Sigma}^2}{2M_{\pi\Sigma}} = \sqrt{|\vec{q}_{K}^{\,\ast}|^2+M_{K}^2}\,,\quad |\vec{q}_{K}^{\,\ast}|=\frac{\sqrt{\lambda(s,M_{\pi\Sigma}^2,M_{K}^2)}}{2M_{\pi\Sigma}}\,.
\end{equation}
In the latter frame, one also finds that
\begin{equation}\label{eq:pSast}
\vec{p}_{\Sigma}^{\,\ast}+\vec{q}_{\pi}^{\,\ast}=\vec{0}\,,\quad  |\vec{p}_{\Sigma}^{\,\ast}| = \frac{\sqrt{\lambda(M_{\pi\Sigma}^2,m_{\Sigma}^2,M_{\pi}^2)}}{2M_{\pi\Sigma}}\,,\quad E_{\Sigma}^{\,\ast} = \frac{M_{\pi\Sigma}^2+m_{\Sigma}^2-M_{\pi}^2}{2M_{\pi\Sigma}}\,,
\end{equation}
and $E_{\pi}^{\,\ast}+E_{\Sigma}^{\,\ast}=M_{\pi\Sigma}$. Energies or momenta without a $\ast$ will always refer to the overall c.m. frame in the following. In both the c.m. and the $\ast$ frame, we choose the photon momentum $\vec{k}$ ($\vec{k}^{\ast}$) to point along the $z-$axis of the respective frame. In App.~\ref{app:kinematics}, we provide the explicit Lorentz transformation that connects the two frames, and a further discussion of the pertaining kinematics.\\
The phase-space integration for the three particles in the final state can be written as
\begin{eqnarray}
  \int\,\mathrm{P.S.} &:=& \int\frac{d^3q_{K}}{(2\pi)^3(2E_{K})}\int\frac{d^3q_{\pi}}{(2\pi)^3(2E_{\pi})}\int\frac{d^3p_{\Sigma}}{(2\pi)^3(2E_{\Sigma})}(2\pi)^4\delta^4(q_{K}+q_{\pi}+p_{\Sigma}-p_{N}-k) \nonumber \\
   &=& \int_{m_{\Sigma}+M_{\pi}}^{\sqrt{s}-M_{K}}dM_{\pi\Sigma}\int d\Omega_{K}\int d\Omega_{\Sigma}^{\,\ast}\,\frac{|\vec{q}_{K}||\vec{p}_{\Sigma}^{\,\ast}|}{8(2\pi)^5\sqrt{s}}\,.\label{eq:PSresult}
\end{eqnarray}
Note that the kaon angles are integrated over in the c.m. frame, while the hyperon angles are integrated in the $\ast$ frame. This representation will prove useful for our purposes (it was also used in \cite{Nakamura:2013boa}). The relativistic flux is given by
\begin{equation}
4\sqrt{(k_{\mu}p_{N}^{\mu})^2-k^2m_{N}^2} = 2(s-m_{N}^2)\,,
\end{equation}
so that the total (unpolarized) cross section is given by
\begin{equation}\label{eq:sigmaTotal}
\sigma_{T}(s) = \frac{1}{2(s-m_{N}^2)}\int\,\mathrm{P.S.}\,\frac{1}{4}\sum_{\sigma,\sigma',\lambda}|\mathcal{M}_{\gamma p\rightarrow\pi K\Sigma}^{\sigma'\sigma\lambda}(s,t_{K},t_{\Sigma},u_{\Sigma},M_{\pi\Sigma})|^2\,,
\end{equation}
where $\mathcal{M}_{\gamma p\rightarrow\pi K\Sigma}^{\sigma'\sigma\lambda}$ is the invariant amplitude for the process, with baryon spinors normalized as \\ $\bar{u}(p,\sigma')u(p,\sigma)=2m\delta_{\sigma'\sigma}$, and the normalization of states $\langle\vec{p}\,'|\vec{p}\,\rangle = (2\pi)^32E_{p}\delta^{3}(\vec{p}\,'-\vec{p})$. Note also the summation over the spins $\sigma,\sigma'$ of the proton and hyperon, and the photon helicities $\lambda$. In some more detail, we write $\mathcal{M}_{\gamma p\rightarrow\pi K\Sigma}^{\sigma'\sigma\lambda}=\bar{u}_{\Sigma}(p_{\Sigma},\sigma')\epsilon^{\lambda}_{\mu}\mathcal{M}^{\mu}u_{p}(p_{N},\sigma)$, where the polarization four-vectors of the photon can be taken as $\epsilon^{\mu}_{\lambda=\pm 1}=(0,\mp 1,-i,0)/\sqrt{2}$\,, and the amplitude $\mathcal{M}^{\mu}$ can be decomposed as  
\begin{eqnarray}
\mathcal{M}^{\mu} &=& \gamma^{\mu}\mathcal{M}_{1} + p_{N}^{\mu}\mathcal{M}_{2} + p_{\Sigma}^{\mu}\mathcal{M}_{3} + q_{K}^{\mu}\mathcal{M}_{4} + k^{\mu}\mathcal{N}_{1} \nonumber \\ &+&  \slashed{k}\left(\gamma^{\mu}\mathcal{M}_{5} + p_{N}^{\mu}\mathcal{M}_{6} + p_{\Sigma}^{\mu}\mathcal{M}_{7} + q_{K}^{\mu}\mathcal{M}_{8} + k^{\mu}\mathcal{N}_{2}\right) \nonumber \\ &+& \slashed{q}_{K}\left(\gamma^{\mu}\mathcal{M}_{9} + p_{N}^{\mu}\mathcal{M}_{10} + p_{\Sigma}^{\mu}\mathcal{M}_{11} + q_{K}^{\mu}\mathcal{M}_{12} + k^{\mu}\mathcal{N}_{3}\right) \nonumber \\ &+&  \slashed{q}_{K}\slashed{k}\left(\gamma^{\mu}\mathcal{M}_{13} + p_{N}^{\mu}\mathcal{M}_{14} + p_{\Sigma}^{\mu}\mathcal{M}_{15} + q_{K}^{\mu}\mathcal{M}_{16} + k^{\mu}\mathcal{N}_{4}\right)\,,\label{eq:MmuDecomp}
\end{eqnarray}
see also \cite{Roberts:1997rz}. The amplitudes $\mathcal{N}_{1-4}$ do not contribute to the photoproduction amplitude, since for real photons $\epsilon_{\mu}k^{\mu}=0$, $k^2=0$, and so we shall suppress them in what follows.
Gauge invariance can be expressed as the set of constraints
\begin{eqnarray}
  (s-m_{N}^2)\mathcal{M}_{2}\, &\overset{!}{=}& \,(u_{\Sigma}-m_{\Sigma}^2)\mathcal{M}_{3} + (t_{K}-M_{K}^2)\mathcal{M}_{4}\,,\nonumber \\
  2\mathcal{M}_{1}+ (s-m_{N}^2)\mathcal{M}_{6}\, &\overset{!}{=}& \,(u_{\Sigma}-m_{\Sigma}^2)\mathcal{M}_{7} + (t_{K}-M_{K}^2)\mathcal{M}_{8}\,,\nonumber \\
  (s-m_{N}^2)\mathcal{M}_{10}\, &\overset{!}{=}& \,(u_{\Sigma}-m_{\Sigma}^2)\mathcal{M}_{11} + (t_{K}-M_{K}^2)\mathcal{M}_{12}\,,\nonumber \\
  2\mathcal{M}_{9}+ (s-m_{N}^2)\mathcal{M}_{14}\, &\overset{!}{=}& \,(u_{\Sigma}-m_{\Sigma}^2)\mathcal{M}_{15} + (t_{K}-M_{K}^2)\mathcal{M}_{16}\,.\label{eq:gaugeinv}
\end{eqnarray}

\section{Low-energy theorems}
\label{sec:LETs}

\begin{figure}[!h]
\centering
\subfigure[\,WT]{\includegraphics[width=0.35\textwidth]{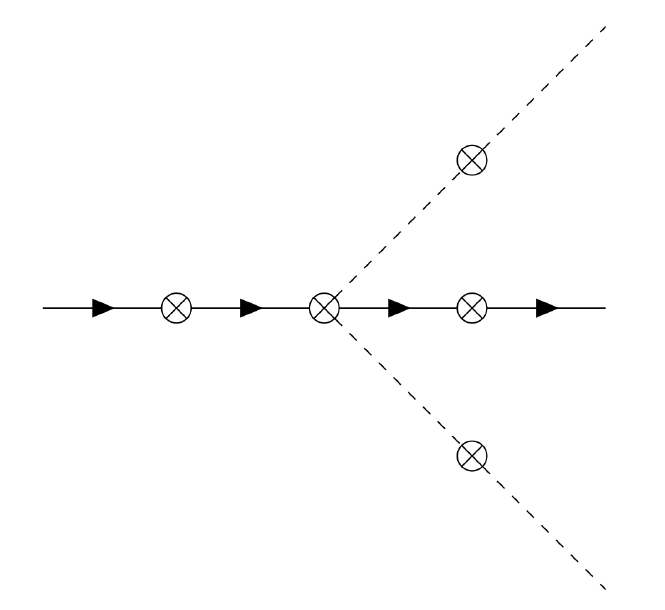}}\\
\subfigure[\,B1, B2]{\includegraphics[width=0.5\textwidth]{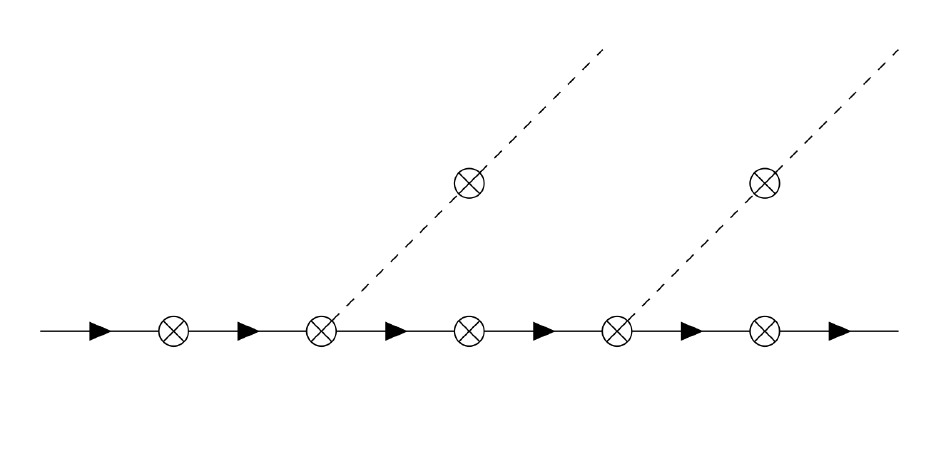}}
\subfigure[\,ANO]{\includegraphics[width=0.31\textwidth]{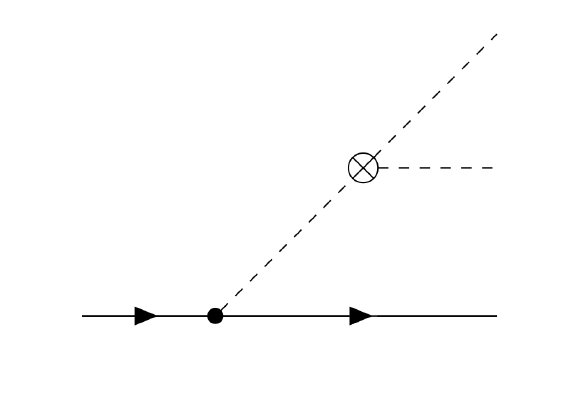}}
\caption{Tree graphs for the two-meson photoproduction amplitude. Directed lines: baryons, dashed lines: pseudoscalar mesons, crossed circles: possible photon insertions. We work only to order $e$, so e.g. the first picture represents five different graphs, with the photon attached to the different crossed vertices, etc. All in all we take into account $5+2\cdot7+1=20$ tree graphs.}
\label{fig:tree_graphs}
\end{figure}

In this section we discuss the predictions of leading-order BChPT for the low-energy behaviour of the cross section. The validity of this approach is certainly limited to the region very close to the overall reaction threshold, where $s\rightarrow s_{thr} := (m_{\Sigma}+M_{\pi}+M_{K})^2$ and $M_{\pi\Sigma}\rightarrow m_{\Sigma}+M_{\pi}$. Taking into account the phase-space factors from Eq.~(\ref{eq:PSresult}), we write in this limit
\begin{equation}\label{eq:d2csthrlim}
\frac{1}{|\vec{q}_{K}||\vec{p}_{\Sigma}^{\,\ast}|}\frac{d^2\sigma}{d\Omega_{K}dM_{\pi\Sigma}} \,\rightarrow\, \frac{\left|\mathcal{A}_{0+,thr}^{1}\right|^2}{(4\pi)^4s_{thr}|\vec{k}|_{thr}}\,,\quad |\vec{k}|_{thr}=\frac{s_{thr}-m_{N}^2}{2\sqrt{s_{thr}}}\,.
\end{equation}
The notation for the amplitude used here anticipates the general notation that will be introduced later in Sec.~\ref{sec:cc_formalism}. We shall use leading-order BChPT to derive expressions for $\mathcal{A}_{0+,thr}^{1}$. The relevant effective Lagrangians are
\begin{eqnarray}
  \mathcal{L}^{(1)}_{MB} &=& i\langle\bar{B}\gamma_{\mu}\lbrack D^{\mu},B\rbrack\rangle - \overset{\circ}{m}\langle\bar{B}B\rangle 
  + \frac{1}{2}D\langle\bar{B}\gamma_{\mu}\gamma_{5}\lbrace u^{\mu},B\rbrace\rangle + \frac{1}{2}F\langle\bar{B}\gamma_{\mu}\gamma_{5}\lbrack u^{\mu},B\rbrack\rangle \,,\label{eq:MB_LagrBMW}\\
  \mathcal{L}^{(2)}_{M} &=& \frac{F_{0}^2}{4}\langle u_{\mu}u^{\mu} + \chi_{+}\rangle\,,\label{eq:MM_LagrBMW}
\end{eqnarray}
where $\overset{\circ}{m}$, $D,F$ and $F_{0}$ are the baryon ground-state octet mass, the baryon axial couplings, and the meson decay constant, respectively, in the three-flavor chiral limit. The baryon fields are collected in the matrix $B$, and the fields of the pseudo-Goldstone-bosons (PGBs, $\pi,K,\eta$) are contained in the matrices $u_{\mu}$, $\chi_{+}$, and the covariant derivative $D^{\mu}$. The notation is further specified in App.~\ref{app:chptnom}. According to the chiral power-counting rules \cite{Gasser:1984gg,Gasser:1987rb}, contributions due to loop graphs are suppressed by powers of small momenta and quark masses at low energies (at least employing a renormalization scheme that preserves the power counting \cite{Ellis:1997kc,Becher:1999he,Fuchs:2003qc}). \\
\quad \\
The classes of tree graphs we take into account are depicted in Fig.~\ref{fig:tree_graphs}.
The first class of graphs, labeled ``WT'', is due to the Weinberg-Tomozawa meson-baryon vertex that stems from the covariant derivative in Eq.~(\ref{eq:MB_LagrBMW}). Classes B1, B2 comprise Born graphs with the axial meson-baryon vertices $\sim D,F$, where in class B1 the pion is emitted first, while in class B2 it is emitted after the kaon. The graph labeled as ``ANO'', strictly speaking, goes beyond the leading-order framework defined by Eqs.~(\ref{eq:MB_LagrBMW}), (\ref{eq:MM_LagrBMW}): it stems from the anomalous Wess-Zumino-Witten Lagrangian \cite{Wess:1971yu,Witten:1983tw}, which is of higher chiral order, and produces a vertex $\gamma K^{+}\rightarrow K^{+}\pi^{0}$. Even though it does not contribute to the threshold limit $\mathcal{A}_{0+,thr}^{1}$, we include it here as an example of a higher-order tree graph. \\
\quad \\
All four classes separately satisfy the gauge-invariance constraints (\ref{eq:gaugeinv}). Analytic results, in the decomposition of Eq.~(\ref{eq:MmuDecomp}), are given in App.~\ref{app:treegraphs}. We note that the graphs of class WT and B1, B2 have also been computed in \cite{Nakamura:2013boa}, employing a truncated expansion in inverse powers of the baryon masses. From the results in App.~\ref{app:treegraphs}, it is a standard procedure to evaluate the cross section via Eqs.~(\ref{eq:sigmaTotal}), (\ref{eq:MmuDecomp}). For the threshold amplitudes $\mathcal{A}_{0+,thr}^{1}$, we then obtain
\begin{eqnarray}
  \mathcal{A}_{0+,thr}^{1}(\gamma p\rightarrow K^{+}\pi^{0}\Sigma^{0}) &=& \frac{e}{4F_{\pi}F_{K}}\left(M_{\pi} + 2(D^2-F^2)(M_{\pi}+M_{K})\right)  + \mathcal{O}(M^2)\,,\label{eq:A0pthrPi0S0}\\
  \mathcal{A}_{0+,thr}^{1}(\gamma p\rightarrow K^{+}\pi^{+}\Sigma^{-}) &=& \frac{2e(M_{\pi}+M_{K})}{3F_{\pi}F_{K}}\left(D^2-3F^2\right) + \mathcal{O}(M^2)\,,\label{eq:A0pthrPipSm}\\
  \mathcal{A}_{0+,thr}^{1}(\gamma p\rightarrow K^{+}\pi^{-}\Sigma^{+}) &=& \frac{e(M_{\pi}+M_{K})}{2F_{\pi}F_{K}}\left(1 + 2F(D-F)+\frac{2}{3}D(D+3F)\right) + \mathcal{O}(M^2) \label{eq:A0pthrPimSp}\,.
\end{eqnarray}
The phases are fixed by the conventions we will discuss later in Sec.~\ref{sec:cc_formalism} and in App.~\ref{app:cc_formalism}. $M$ stands collectively for PGB masses $M_{\pi},\,M_{K}$ or $M_{\eta}$. Note that baryon mass differences and $F_{\pi,K}-F_{0}$ are booked as $\mathcal{O}(M^2)$ and $\mathcal{O}(M^2\log M)$ according to the chiral counting rules. \\
With $e=+\sqrt{4\pi\alpha_{\mathrm{QED}}}$, $D\approx 0.8$, $F\approx 0.5$, we arrive at a first, very rough, order-of-magnitude estimate,
\begin{displaymath}
\left|\mathcal{A}_{0+,thr}^{1}(\gamma p\rightarrow K^{+}\pi\Sigma)\right| \sim \frac{e(M_{\pi}+M_{K})}{2F_{\pi}F_{K}} \sim 10\,\mathrm{GeV}^{-1}\,.
\end{displaymath}
Our formulae are flexible enough to allow for other $S=-1$ meson-baryon states than $\pi\Sigma$ to be produced with the $K^{+}$. This is important, because we intend to use the photoproduction amplitudes as a kernel in a coupled-channel formalism later on. Replacing the masses, charges and coupling factors at the appropriate places, we obtain further results for threshold amplitudes:
\begin{eqnarray}
  \mathcal{A}_{0+,thr}^{1}(\gamma p\rightarrow K^{+}\pi^{0}\Lambda) &=& \frac{e}{4\sqrt{3}F_{\pi}F_{K}}\left(3M_{\pi} -2(D+F)(D+3F)(M_{\pi}+M_{K})\right)  + \mathcal{O}(M^2),\label{eq:A0pthrPi0L0}\,\,\\
  \mathcal{A}_{0+,thr}^{1}(\gamma p\rightarrow K^{+}K^{-}p) &=& \frac{2eM_{K}}{3F_{K}^2}\left(3-2D^2-6F^2\right) + \mathcal{O}(M^2),\label{eq:A0pthrKmp}\\
  \mathcal{A}_{0+,thr}^{1}(\gamma p\rightarrow K^{+}\bar{K}^{0}n) &=& \frac{eM_{K}}{2F_{K}^2} + \mathcal{O}(M^2)\,.\label{eq:A0pthrKbar0n}
\end{eqnarray}
Even though the low-energy limits in Eqs.~(\ref{eq:d2csthrlim}), (\ref{eq:A0pthrPi0S0})-(\ref{eq:A0pthrKbar0n}) are rigorous results from BChPT, we do not expect them to be directly useful for a comparison to data. Higher-order corrections to the above results are probably large, not only because the expansion parameters $M_{K}$ and $M_{\eta}$ are not very small compared to a typical hadronic scale of $\sim 1\,\mathrm{GeV}$, but mainly due to the presence of the $\Lambda^{\ast}(1405)$ resonance close to the $\bar{K}N$ threshold. This resonant enhancement clearly represents a non-perturbative feature of the $S=-1$ meson-baryon scattering amplitude in the low-energy region, which cannot be described by the EFT in its standard formulation. The above low-energy theorems can still be of use, however, as {\em theoretical constraints\,} for {\em models\,} describing the $K^{+}MB$ photoproduction process. Moreover, the tree graph results we present here can constitute the leading terms in a photoproduction kernel for a non-perturbative framework, which captures the effects due to the  $\Lambda^{\ast}$ resonance. We will outline such a framework in Sec.~\ref{sec:cc_formalism}. Before we do this, it will be instructive to assess the impact of the presence of a $\Lambda^{\ast}$ resonance from a different, greatly simplified point of view. 

\clearpage

\section{Two-step model with an explicit $\mathbf{\Lambda^{\ast}}$ resonance}
\label{sec:twostep}

\begin{figure}[h!]
\centering
\includegraphics[width=0.5\textwidth]{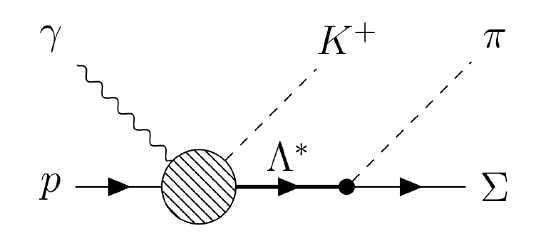}
\caption{Resonance exchange graph for the two-step approximation. The blob on the left contains the $K^{+}\Lambda^{\ast}$ photoproduction amplitude.}
\label{fig:twostep}
\end{figure}

\begin{figure}[h!]
\centering
\includegraphics[width=0.6\textwidth]{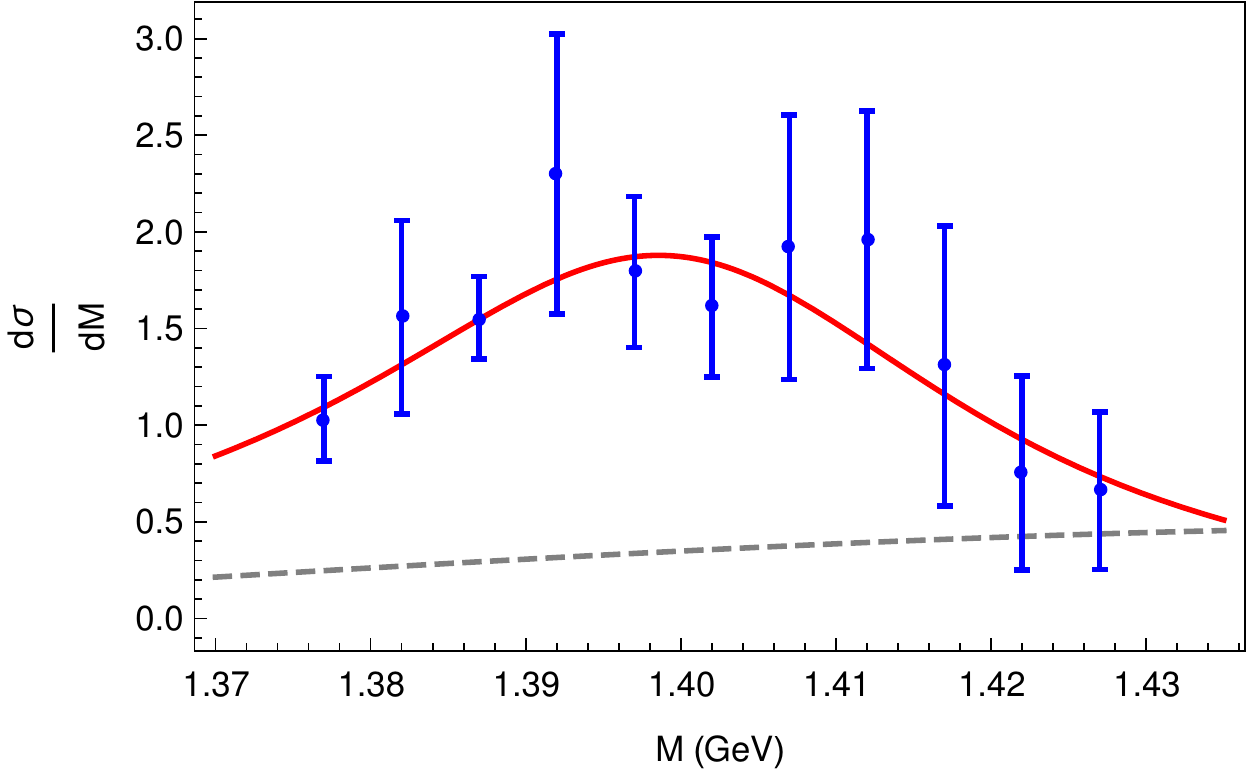}
\caption{$\frac{d\sigma}{dM_{\pi\Sigma}}$ for $\gamma p\rightarrow K^{+}\pi^{0}\Sigma^{0}$ at $s=(2\,\mathrm{GeV})^2$, in $\mu\mathrm{b}/\mathrm{GeV}$. Red curve: fit with Eq.~(\ref{eq:BWcrosssecintK}). Gray dashed curve: contribution from WT\,+\,Born tree graphs. Data from \cite{Moriya:2013eb}.}
\label{fig:BWplot}
\end{figure}

As we will be mainly concerned with the extraction of the $\Lambda^{\ast}$ properties from $\pi\Sigma$ invariant mass distributions $d\sigma/dM_{\pi\Sigma}$, it seems worthwhile to consider a simplified hadronic model, illustrated in Fig.~\ref{fig:twostep}, where the photoproduction process is assumed to take place in two separable steps: first, the $\Lambda^{\ast}$ is produced together with a $K^{+}$, and subsequently decays into a $\pi\Sigma$ state. We invoke on-shell factorization here, so that the blob on the left-hand side of Fig.~\ref{fig:twostep} stands for an on-shell $\Lambda^{\ast}$ photoproduction amplitude $\mathcal{S}_{\Lambda^{\ast}}^{\mu}$, while the strength of the $\Lambda^{\ast}\pi\Sigma$ vertex on the right-hand side can be related to the width of the $\Lambda^{\ast}$, $\Gamma_{\Lambda^{\ast}}$. The rationale behind this is that we are now interested mainly in the resonance pole terms in the variable $M_{\pi\Sigma}$, so that we can neglect all kinds of background contributions. In analogy to Eq.~(\ref{eq:MmuDecomp}), we can decompose (leaving out structures $\sim k^{\mu}$)
\begin{equation}\label{eq:SmuDecomp}
\mathcal{S}_{\Lambda^{\ast}}^{\mu} = \mathcal{S}_{1}\gamma^{\mu} + \mathcal{S}_{2}p_{N}^{\mu} + \mathcal{S}_{3}q_{K}^{\mu} + \mathcal{S}_{4}\slashed{k}\gamma^{\mu}+\mathcal{S}_{5}\slashed{k}p_{N}^{\mu} + \mathcal{S}_{6}\slashed{k}q_{K}^{\mu}\,,
\end{equation}
wherein the structure functions $\mathcal{S}_{i}$ depend on $s=(p_{N}+k)^2$ and $t_{K}=(q_{K}-k)^2$. The kinematics are again chosen as detailed in Sec.~\ref{sec:generalities} and App.~\ref{app:kinematics}. In particular, in the c.m. frame where $\vec{p}_{N}+\vec{k}=\vec{0}$, the photon momentum $\vec{k}$ points along the $z$-direction, and $\vec{q}_{K}$ lies in the $x-z$ plane, forming an angle $\theta_{K}$ with $\vec{k}$. We can eliminate $\mathcal{S}_{3}$ and $\mathcal{S}_{6}$ via gauge-invariance constraints, and define the combinations
\begin{eqnarray*}
  \mathcal{B}^{1}_{\Lambda^{\ast}} &=& \sqrt{E_{\Lambda^{\ast}}+m_{\Lambda^{\ast}}}\left(\mathcal{S}_{1}+(\sqrt{s}+m_{N})\mathcal{S}_{4}\right)\sqrt{E_{N}-m_{N}}\,/(8\pi\sqrt{s})\,,\\
  \mathcal{B}^{2}_{\Lambda^{\ast}} &=& \sqrt{E_{\Lambda^{\ast}}-m_{\Lambda^{\ast}}}\left(\mathcal{S}_{1}-(\sqrt{s}-m_{N})\mathcal{S}_{4}\right)\sqrt{E_{N}+m_{N}}\,/(8\pi\sqrt{s})\,,\\
  \mathcal{B}^{3}_{\Lambda^{\ast}} &=& \sqrt{E_{\Lambda^{\ast}}+m_{\Lambda^{\ast}}}\left(\mathcal{S}_{1}+\frac{1}{2}(\sqrt{s}+m_{N})\mathcal{S}_{2} + \frac{1}{2}(s-m_{N}^2)\mathcal{S}_{5}\right)\sqrt{E_{N}-m_{N}}\,/(8\pi\sqrt{s})\,,\\
  \mathcal{B}^{4}_{\Lambda^{\ast}} &=& \sqrt{E_{\Lambda^{\ast}}-m_{\Lambda^{\ast}}}\left(\mathcal{S}_{1}-\frac{1}{2}(\sqrt{s}-m_{N})\mathcal{S}_{2} + \frac{1}{2}(s-m_{N}^2)\mathcal{S}_{5}\right)\sqrt{E_{N}+m_{N}}\,/(8\pi\sqrt{s})\,,
\end{eqnarray*}
where $E_{\Lambda^{\ast}}$ and $E_{N}$ are the c.m. \hspace{-0.15cm}energies of the outgoing $\Lambda^{\ast}$ and the incoming proton. These combinations are convenient for a low-energy analysis, because the leading multipole amplitude $\mathcal{M}_{0+}^{\Lambda^{\ast}}$ (the analogue of the electric dipole amplitude $\mathcal{E}_{0+}^{\Lambda}$ in $K^{+}\Lambda$ photoproduction) is only contained in $\mathcal{B}^{1}_{\Lambda^{\ast}}$, and $\int\frac{d\Omega_{K}}{4\pi}\mathcal{B}^{1}_{\Lambda^{\ast}}$ reduces to $\mathcal{M}_{0+}^{\Lambda^{\ast}}$ (up to $\ell>1$ contributions). Furthermore, $\mathcal{E}_{1-}^{\Lambda^{\ast}}$ (the analogue of $\mathcal{M}_{1-}^{\Lambda}$) is only contained in $\mathcal{B}^{2}_{\Lambda^{\ast}}$, $\mathcal{B}^{4}_{\Lambda^{\ast}}$ contains no s- or p-wave multipoles at all, and the $\mathcal{B}^{i}_{\Lambda^{\ast}}$ have no ``kaon pole'' at $t_{K}=M_{K}^2$. With the help of the above combinations, we can write the differential cross section for $\Lambda^{\ast}$ photoproduction in a ``manifestly positive'' way ($z_{K}:=\cos\theta_{K}$),
\begin{eqnarray}
\frac{d\sigma}{d\Omega}\biggr|_{\gamma p\rightarrow K^{+}\Lambda^{\ast}} &=& \frac{|\vec{q}_{K}|}{|\,\vec{k}\,|}\left|\mathcal{B}_{\Lambda^{\ast}}\right|^2\,,\label{eq:dcsLstar}\\
4\left|\mathcal{B}_{\Lambda^{\ast}}\right|^2 &=& (1-z_{K})\left|\mathcal{B}^{1}_{\Lambda^{\ast}}+\mathcal{B}^{2}_{\Lambda^{\ast}}\right|^2 + (1+z_{K})\left|\mathcal{B}^{1}_{\Lambda^{\ast}} -\mathcal{B}^{2}_{\Lambda^{\ast}}\right|^2 \nonumber \\ &+& (1-z_{K})\left|\mathcal{B}^{1}_{\Lambda^{\ast}}+\mathcal{B}^{2}_{\Lambda^{\ast}}+\frac{2|\vec{q}_{K}|(1+z_{K})}{M_{K}^2-t_{K}}\left((\sqrt{s}+m_{N})\mathcal{B}^{3}_{\Lambda^{\ast}}+(\sqrt{s}-m_{N})\mathcal{B}^{4}_{\Lambda^{\ast}}\right)\right|^2 \nonumber \\
    &+& (1+z_{K})\left|\mathcal{B}^{1}_{\Lambda^{\ast}}-\mathcal{B}^{2}_{\Lambda^{\ast}}-\frac{2|\vec{q}_{K}|(1-z_{K})}{M_{K}^2-t_{K}}\left((\sqrt{s}+m_{N})\mathcal{B}^{3}_{\Lambda^{\ast}}-(\sqrt{s}-m_{N})\mathcal{B}^{4}_{\Lambda^{\ast}}\right)\right|^2 \,.\nonumber
\end{eqnarray}
We will refer to this form for the $\Lambda^{\ast}$ photoproduction cross section in the next section. Meanwhile, instead of discussing any specific model for $\mathcal{S}_{\Lambda^{\ast}}^{\mu}$, we just note that it is basically multiplied by the resonance propagator and the decay vertex to give the resonance pole contribution to $\mathcal{M}^{\mu}(\gamma p\rightarrow K^{+}\pi\Sigma)$. Studying these contributions, one observes that the strengths (residues) of those pole terms only depend on the Mandelstam variables $s$ and $t_{K}$, in other words, they do not depend on the direction of the outgoing $\Sigma$ hyperon (specified by angles $\theta_{\Sigma}^{\ast},\,\phi_{\Sigma}^{\ast}$ in spherical coordinates), as one can expect for the contribution due to an s-wave resonance. Moreover, such an angular dependence would enter through $p_{\Sigma}^{\mu}$ and therefore always comes with powers of $|\vec{p}_{\Sigma}^{\,\ast}|$, which are suppressed at low energies. In relation to this, it is also worth noting that the structures $\sim p_{\Sigma}^{\mu}$ (i.e. $\mathcal{M}_{3,7,11,15}$) do not receive $\Lambda^{\ast}$ pole term contributions. \\
The considerations of this section motivate a kind of Breit-Wigner approximation for the $\pi\Sigma$ invariant mass distribution, which we write in the form
\begin{equation}\label{eq:BWcrosssecintK}
\frac{d\sigma}{dM_{\pi\Sigma}} \approx \frac{|\vec{q}_{K}|}{|\,\vec{k}\,|}\left(\frac{2}{3}\Gamma_{\Lambda^{\ast}}\right)\biggl|\frac{2M_{\pi\Sigma}}{M_{\pi\Sigma}^2-m_{\Lambda^{\ast}}^2+iM_{\pi\Sigma}\Gamma_{\Lambda^{\ast}}}\biggr|^2\left|\overline{\mathcal{B}}_{\Lambda^{\ast}}(s)\right|^2\,,
\end{equation}
where $\left|\overline{\mathcal{B}}_{\Lambda^{\ast}}(s)\right|$ is merely introduced as a fit parameter for fixed $s$. A simple tentative fit to a dozen of data points for the invariant mass distribution for $\gamma p\rightarrow K^{+}\pi^{0}\Sigma^{0}$ at $\sqrt{s}=2\,\mathrm{GeV}$ around the resonance peak (see Fig.~\ref{fig:BWplot}) returns
\begin{displaymath}
\left|\overline{\mathcal{B}}_{\Lambda^{\ast}}(s=(2\,\mathrm{GeV})^2)\right|\approx 0.0155\,\mathrm{GeV}^{-1}, \quad m_{\Lambda^{\ast}}\approx 1400\,\mathrm{MeV},\quad \Gamma_{\Lambda^{\ast}}\approx 49\,\mathrm{MeV}\,.
\end{displaymath}

\quad \\
It is, of course, well known \cite{Hemingway:1984pz,Dalitz:1991sq,Oller:2000fj} that simple Breit-Wigner approximations are insufficient to understand all aspects of $\Lambda^{\ast}(1405)$ physics, and thus one should not take such a fit to a small selection of data points very seriously. It is presented here merely to demonstrate the dominance of the effects due to the $\Lambda^{\ast}$ pole(s) in the $\pi\Sigma$ invariant-mass distribution. The current state-of-the-art description of the $\Lambda^{\ast}(1405)$ utilizes coupled-channel meson-baryon scattering amplitudes, with kernels derived from chiral Lagrangians as (\ref{eq:MB_LagrBMW}) and its higher-order extensions. Based on the observations made in the present section, we will propose, in the next section, a formalism that allows to incorporate such coupled-channel amplitudes in the description of the $K^{+}\pi\Sigma$ photoproduction process in the $\Lambda^{\ast}(1405)$ region.

\newpage

\section{Coupled-channel formalism}
\label{sec:cc_formalism}

\begin{figure}[h!]
\centering
\includegraphics[width=0.5\textwidth]{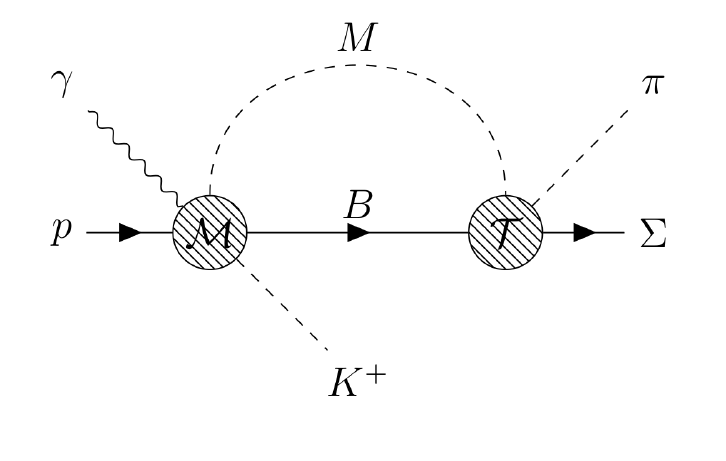}
\caption{Representation of the final-state interaction of the $S=-1$ meson-baryon pair $MB$. $\mathcal{M}$ is the amplitude for $\gamma p\rightarrow K^{+}MB$, and $\mathcal{T}$ is the $S=-1$ meson-baryon scattering amplitude, which can be decomposed into partial waves $f_{\ell\pm}$ \cite{Chew:1957zz}.}
\label{fig:MBFSI}
\end{figure}

In the coupled-channel formalism of ``Unitarized ChPT'', one constructs meson-baryon partial-wave scattering amplitudes $f_{\ell\pm}^{c',c}(M_{\pi\Sigma})$, which aim to describe the scattering from channel $c$ to channel $c'$ (here, $c,c'=\pi\Sigma,\,\bar{K}N,\,\eta\Lambda,\ldots$) for angular momentum $J=\ell\pm\frac{1}{2}$. These amplitudes are designed such that they are consistent with ChPT up to some fixed order of the low-energy expansion (in practice, usually on tree level, i.e. $\mathcal{O}(p)$ or $\mathcal{O}(p^2)$), and, at the same time, conform with the requirement of ``coupled-channel unitarity'',
\begin{equation}\label{eq:unif}
\mathrm{Im}(f_{\ell\pm}) = (f_{\ell\pm})^{\dagger}(|\vec{p}^{\,\ast}|)(f_{\ell\pm})
\end{equation}
above the lowest reaction threshold. Here, $(f_{\ell\pm})$ denotes a matrix in the space of the considered meson-baryon channels $c,c'$, with entries $f_{\ell\pm}^{c',c}(M_{\pi\Sigma})$, while $(|\vec{p}^{\,\ast}|)$ is a diagonal matrix in this space, the entries of which are given by the moduli of the three-momenta (for each channel $c$) in the meson-baryon c.m. frame (in the context of the present paper, this agrees with our $\ast$ frame, where $M_{\pi\Sigma}$ is the c.m. energy of the meson-baryon pair - see also Eq.~(\ref{eq:pSast})) above the threshold of channel $c$, and zero below it. \\
\quad \\
It is our aim to implement the scattering amplitudes $f_{0+}$ in the photoproduction formalism, in such a way that it describes the final-state s-wave interaction of the $S=-1$ meson-baryon pair $MB$ produced in $\gamma p\rightarrow K^{+}MB$ (compare Fig.~\ref{fig:MBFSI}). Clearly, we have to find the combinations of (partial-wave projections of) the structure functions $\mathcal{M}_{i=1,\ldots,16}$ which project on the $\ell=0$ state of this meson-baryon pair. Remembering the considerations of the previous section, we might expect that there are four such combinations, as there are four independent structure functions (e.g., $\mathcal{S}_{1,2,4,5}$) in the $\Lambda^{\ast}$ photoproduction amplitude after imposing gauge invariance. Moreover, remembering the four so-called CGLN amplitudes $\mathcal{F}_{1,\ldots,4}$ for pion photoproduction $\gamma N\rightarrow \pi N$ \cite{Chew:1957tf}, there should be the same number of projections for $\ell=1$, $J=\frac{1}{2}$. \\
In fact, one can construct amplitudes $\mathcal{A}^{i=1,\ldots,4}_{0+}(s,M_{\pi\Sigma}^2,t_{K})$, $\mathcal{A}^{i=1,\ldots,4}_{1-}(s,M_{\pi\Sigma}^2,t_{K})$ from the $\mathcal{M}_{i}$ which have simple unitarity relations with the pertinent partial-wave scattering amplitudes,
\begin{equation}\label{eq:uniA}
\mathrm{Im}(\mathcal{A}^{i}_{\ell\pm}) = (f_{\ell\pm})^{\dagger}(|\vec{p}^{\,\ast}|)(\mathcal{A}^{i}_{\ell\pm})\,,\quad i=1,\ldots,4,\quad \ell\pm = 0+,\,1-\,.
\end{equation}
These can be found e.g. by applying the Cutkosky rules \cite{Cutkosky:1960sp} to the $MB$ loop in Fig.~\ref{fig:MBFSI}, and studying the consequences of unitarity for the various invariant structures in $\mathcal{M}^{\mu}$, or equivalently by the methods used in \cite{Chew:1957tf,Pearlstein:1957zz,Berends:1967vi} to extract the multipole amplitudes for single-meson photoproduction. Explicit expressions are provided in Appendix \ref{app:cc_formalism}. Even though we are mainly interested in the s-wave final state interaction, we also give the ``$1-$'' amplitudes, because in the analysis they practically come ``for free'' together with the results for the ``$0+$'' sector. This analysis is somewhat simplified employing an approximation which is motivated by the observations made in the previous section: confining ourselves to a low-energy analysis in the $\Lambda^{\ast}(1405)$ energy region, it is reasonable to neglect higher partial waves $\sim\mathcal{Y}_{\ell>0, m}(\theta_{\Sigma}^{\ast},\phi_{\Sigma}^{\ast})$ in the decomposition of the $\mathcal{M}_{i}$ into spherical harmonics, and express the projections solely through
\begin{equation}\label{eq:MbarApprox}
\overline{\mathcal{M}}_{i}(s,M_{\pi\Sigma}^2,t_{K}):=\int\frac{d\Omega_{\Sigma}^{\ast}}{4\pi}\mathcal{M}_{i}(s,M_{\pi\Sigma}^2,t_{K},t_{\Sigma},u_{\Sigma})\,. 
\end{equation}
We have seen in Sec.~\ref{sec:twostep} that s-wave resonance pole terms in the photoproduction amplitude are not affected by this approximation. We also note that this approximation can be used for any c.m. energy $\sqrt{s}$, as long as $M_{\pi\Sigma}$ stays sufficiently close to the $MB$ threshold region. Therefore, kinematics involving high-energy kaons can, in principle, be treated within the framework of this section. Of course, the BChPT tree graphs might not be sufficient for this purpose, and the elementary photoproduction amplitude would have to be amended, but that is another matter. \\
\quad \\
Neglecting contributions due to $\pi\Sigma$ states with $J>\frac{1}{2}$, the double-differential cross section for $\gamma p\rightarrow K^{+}\pi\Sigma$ can be expressed through the $\mathcal{A}^{i}_{\ell\pm}$ amplitudes as follows,
\begin{eqnarray}
  \frac{d^2\sigma}{d\Omega_{K}dM_{\pi\Sigma}} &=& \frac{|\vec{q}_{K}||\vec{p}_{\Sigma}^{\,\ast}|}{(4\pi)^4s|\vec{k}|}|\mathcal{A}|^2\,,\label{eq:d2csA}\\
  4|\mathcal{A}|^2 &=& (1-z_{K})\left|\mathcal{A}_{0+}^{1} + \mathcal{A}_{0+}^{2}\right|^2  +  (1+z_{K})\left|\mathcal{A}_{0+}^{1} - \mathcal{A}_{0+}^{2}\right|^2 \nonumber \\
    &+& (1-z_{K})\biggl|\mathcal{A}_{0+}^{1} + \mathcal{A}_{0+}^{2}  + \frac{2|\vec{q}_{K}|(1+z_{K})}{M_{K}^2-t_{K}}\left((\sqrt{s}+m_{N})\mathcal{A}_{0+}^{3} + (\sqrt{s}-m_{N})\mathcal{A}_{0+}^{4}\right)\biggr|^2 \nonumber \\  &+& (1+z_{K})\biggl|\mathcal{A}_{0+}^{1} - \mathcal{A}_{0+}^{2}  - \frac{2|\vec{q}_{K}|(1-z_{K})}{M_{K}^2-t_{K}}\left((\sqrt{s}+m_{N})\mathcal{A}_{0+}^{3} - (\sqrt{s}-m_{N})\mathcal{A}_{0+}^{4}\right)\biggr|^2 \nonumber \\
&+& (1-z_{K})\left|\mathcal{A}_{1-}^{1} + \mathcal{A}_{1-}^{2}\right|^2  +  (1+z_{K})\left|\mathcal{A}_{1-}^{1} - \mathcal{A}_{1-}^{2}\right|^2 \nonumber \\
    &+& (1-z_{K})\biggl|\mathcal{A}_{1-}^{1} + \mathcal{A}_{1-}^{2}  + \frac{2|\vec{q}_{K}|(1+z_{K})}{M_{K}^2-t_{K}}\left((\sqrt{s}-m_{N})\mathcal{A}_{1-}^{3} + (\sqrt{s}+m_{N})\mathcal{A}_{1-}^{4}\right)\biggr|^2 \nonumber \\  &+& (1+z_{K})\biggl|\mathcal{A}_{1-}^{1} - \mathcal{A}_{1-}^{2}  - \frac{2|\vec{q}_{K}|(1-z_{K})}{M_{K}^2-t_{K}}\left((\sqrt{s}-m_{N})\mathcal{A}_{1-}^{3} - (\sqrt{s}+m_{N})\mathcal{A}_{1-}^{4}\right)\biggr|^2\,,\nonumber 
\end{eqnarray}
where $z_{K}\equiv\cos\theta_{K}$. The reader should note the formal similarity of the terms involving the $\mathcal{A}_{0+}^{i}$ to the expression in Eq.~(\ref{eq:dcsLstar}). Indeed, computing amplitudes from graphs like the one in Fig.~\ref{fig:twostep}, one finds that the $\Lambda^{\ast}$ pole terms in the resulting $\mathcal{A}_{0+}^{i}$ are proportional to the combinations $\mathcal{B}^{i}_{\Lambda^{\ast}}$. In the same way, one could relate the terms involving the  $\mathcal{A}_{1-}^{i}$ to pole terms due to a $S=-1$, spin $\frac{1}{2}$ p-wave resonance.\\
 Approaching the reaction threshold, only the combination $\mathcal{A}_{0+}^{1}$ survives in (\ref{eq:d2csA}), which explains the notation used earlier in Eq.~(\ref{eq:d2csthrlim}).  \\
 \quad \\
Under the additional assumption that we can neglect the $\mathcal{A}_{1-}^{i}$ and the $\theta_{K}$-dependence of the $\mathcal{A}_{0+}^{i}$, we can perform the integration over the kaon angles analytically, to obtain
\begin{eqnarray}\label{eq:dsigmadMsimple}
\frac{d\sigma}{dM_{\pi\Sigma}} &=& \frac{|\vec{q}_{K}||\vec{p}_{\Sigma}^{\,\ast}|}{(4\pi)^3s|\vec{k}|}\biggl(\left|\mathcal{A}_{0+}^{1}\right|^2 + \left|\mathcal{A}_{0+}^{2}\right|^2 + \left|\frac{2\sqrt{s}\mathcal{A}_{0+}^{3}}{\sqrt{s}-m_{N}}\right|^2\left(\frac{E_{K}}{|\vec{q}_{K}|}\mathrm{artanh}\left(\frac{|\vec{q}_{K}|}{E_{K}}\right)-1\right) \nonumber \\ & & \hspace{4.96cm} +\, \left|\frac{2\sqrt{s}\mathcal{A}_{0+}^{4}}{\sqrt{s}+m_{N}}\right|^2\left(\frac{E_{K}}{|\vec{q}_{K}|}\mathrm{artanh}\left(\frac{|\vec{q}_{K}|}{E_{K}}\right)-1\right) \nonumber \\ & & \hspace{2cm} +\,\frac{2\sqrt{s}}{\sqrt{s}+m_{N}}\left(\frac{E_{K}}{|\vec{q}_{K}|} - \frac{M_{K}^2}{|\vec{q}_{K}|^2}\mathrm{artanh}\left(\frac{|\vec{q}_{K}|}{E_{K}}\right)\right)\mathrm{Re}(\mathcal{A}_{0+}^{1}\mathcal{\bar{A}}_{0+}^{4}) \nonumber \\ & & \hspace{2cm} +\,\frac{2\sqrt{s}}{\sqrt{s}-m_{N}}\left(\frac{E_{K}}{|\vec{q}_{K}|} - \frac{M_{K}^2}{|\vec{q}_{K}|^2}\mathrm{artanh}\left(\frac{|\vec{q}_{K}|}{E_{K}}\right)\right)\mathrm{Re}(\mathcal{A}_{0+}^{2}\mathcal{\bar{A}}_{0+}^{3})  \nonumber \\ & & \hspace{2cm} +\, \frac{4s}{s-m_{N}^2}\left(\frac{2E_{K}^2+M_{K}^2}{|\vec{q}_{K}|^2}\mathrm{artanh}\left(\frac{|\vec{q}_{K}|}{E_{K}}\right) - \frac{3E_{K}}{|\vec{q}_{K}|}\right)\mathrm{Re}(\mathcal{A}_{0+}^{3}\mathcal{\bar{A}}_{0+}^{4}) \biggr)\,.
\end{eqnarray}
In general, however, the simple formula (\ref{eq:dsigmadMsimple}) will not apply, and one has to do the integration over the kaon angles numerically from (\ref{eq:d2csA}). \\
\quad\\
Returning to the partial-wave unitarity statement in Eq.~(\ref{eq:uniA}), the coupled-channel formalism we propose is now easily explained: We have to construct an elementary photoproduction amplitude, e.g. from the tree graphs of App.~\ref{app:treegraphs}, and compute the according projected amplitudes $\mathcal{A}^{i(\mathrm{tree})}_{\ell\pm}$, $i=1,\ldots,4$, as detailed here and in App.~\ref{app:cc_formalism}. The set of ``unitarized'' amplitudes for $\gamma p\rightarrow K^{+}MB$ will then be taken as the coupled-channel vector
\begin{equation}\label{eq:modelA}
(\mathcal{A}^{i}_{\ell\pm}) = (\mathcal{A}^{i(\mathrm{tree})}_{\ell\pm}) + (f_{\ell\pm})\left(8\pi M_{\pi\Sigma}G(M_{\pi\Sigma})\right)(\mathcal{A}^{i(\mathrm{tree})}_{\ell\pm})\,,\quad i=1,\ldots,4,\quad \ell\pm = 0+,\,1-\,,
\end{equation}
where $G(M_{\pi\Sigma})$ is a diagonal channel-space matrix, with entries given by suitably regularized loop integrals
\begin{equation}
G^{c=MB}(M_{\pi\Sigma}) = \frac{1}{i}\int_{\mathrm{reg.}}\frac{d^{4}l}{(2\pi)^4}\frac{1}{((p_{\Sigma}+q_{\pi}-l)^2-m_{B}^2+i\epsilon)(l^2-M_{M}^2+i\epsilon)}\,.
\end{equation}
The $MB=\pi\Sigma$ entries of the vector $(\mathcal{A}^{i}_{\ell\pm})$ can then be inserted in formula (\ref{eq:d2csA}) to obtain the according cross sections. Using the fact that
\begin{equation}
  8\pi M_{\pi\Sigma}\mathrm{Im}\,G^{MB}(M_{\pi\Sigma})=|\vec{p}_{B}^{\,\ast}|\Theta(M_{\pi\Sigma}-(m_{B}+M_{M}))\,,
\end{equation}
where $\Theta(\cdot)$ denotes the Heaviside step function, together with Eq.~(\ref{eq:unif}), it is straightforward to show that the ansatz (\ref{eq:modelA}) solves the unitarity requirement (\ref{eq:uniA}). \\
\quad \\
Could there be a conflict between Eqs.~(\ref{eq:d2csA})-(\ref{eq:modelA}) and gauge invariance? The question is non-trivial: Even a gauge-invariant tree-level amplitude $\mathcal{M}^{\mu}_{(\mathrm{tree})}$ does not necessarily generate a gauge-invariant ``unitarized'' amplitude when it is just plugged into a loop integral to couple it to the final-state interaction, since it will usually depend on the loop momentum that is integrated over (e.g., the $\mathcal{M}_{3,7,11,15}$-structures would come with the baryon loop four-momentum). We refer here to the discussions in \cite{Borasoy:2005zg,Borasoy:2007ku,Mai:2012wy}. In our present framework, the issue is resolved as follows: given the functions $\mathcal{A}^{i}_{\ell\pm}(s,M_{\pi\Sigma}^2,t_{K})$, we can find a set of invariant amplitudes $\mathcal{M}_{i}^{\mathcal{C}}$ (as given in App.~\ref{app:cc_formalism}) which form a gauge-invariant amplitude by construction. The cross-section calculated directly from this gauge-invariant amplitude exactly equals the one in Eq.~(\ref{eq:d2csA}). The difference between the corresponding invariant amplitudes $\mathcal{M}_{i}^{\mathcal{C}}$ and $\mathcal{M}_{i}$ thus resides in the higher meson-baryon partial waves, which are neglected here anyway.\\
This strategy works for every gauge-invariant set of $\mathcal{A}^{i(\mathrm{tree})}_{\ell\pm}$ in (\ref{eq:modelA}), which allows for some flexibility in the construction of models for these functions. For example, we could add higher-order contact terms to the ChPT tree graphs, etc. All these models will yield cross-sections that are in accord with s-wave coupled-channel unitarity, gauge invariance and the chiral low-energy theorems (as long as the model used for the $f_{\ell\pm}$ does not spoil the proper low-energy behaviour) at the same time.\\

To give a simple example for the entries in $(\mathcal{A}^{i(\mathrm{tree})}_{\ell\pm})$ , we obtain the following contributions from the WT class tree graphs, for the reaction $\gamma p\rightarrow K^{+}\pi^{0}\Sigma^{0}$ ($E_{\pi\Sigma}:=\sqrt{s}-E_{K}$):
\begin{eqnarray*}
  \mathcal{A}_{0+}^{1,\mathrm{WT}} &=& \frac{e}{8F_{\pi}F_{K}}\sqrt{\frac{E_{\Sigma}^{\ast}+m_{\Sigma}}{2M_{\pi\Sigma}}}\sqrt{\frac{E_{\pi\Sigma}+M_{\pi\Sigma}}{2\sqrt{s}}}\left(2M_{\pi\Sigma}-m_{\Sigma}-m_{N}\right) = \mathcal{A}_{0+}^{3,\mathrm{WT}}\,,\\
  \mathcal{A}_{0+}^{2,\mathrm{WT}} &=& -\frac{e}{8F_{\pi}F_{K}}\sqrt{\frac{E_{\Sigma}^{\ast}+m_{\Sigma}}{2M_{\pi\Sigma}}}\sqrt{\frac{E_{\pi\Sigma}-M_{\pi\Sigma}}{2\sqrt{s}}}\left(2M_{\pi\Sigma}-m_{\Sigma}-m_{N}\right) = \mathcal{A}_{0+}^{4,\mathrm{WT}}\,,\\
  \mathcal{A}_{1-}^{1,\mathrm{WT}} &=& \frac{e}{8F_{\pi}F_{K}}\sqrt{\frac{E_{\Sigma}^{\ast}-m_{\Sigma}}{2M_{\pi\Sigma}}}\sqrt{\frac{E_{\pi\Sigma}+M_{\pi\Sigma}}{2\sqrt{s}}}\left(2M_{\pi\Sigma}+m_{\Sigma}+m_{N}\right) = \mathcal{A}_{1-}^{3,\mathrm{WT}}\,,\\
  \mathcal{A}_{1-}^{2,\mathrm{WT}} &=& -\frac{e}{8F_{\pi}F_{K}}\sqrt{\frac{E_{\Sigma}^{\ast}-m_{\Sigma}}{2M_{\pi\Sigma}}}\sqrt{\frac{E_{\pi\Sigma}-M_{\pi\Sigma}}{2\sqrt{s}}}\left(2M_{\pi\Sigma}+m_{\Sigma}+m_{N}\right) = \mathcal{A}_{1-}^{4,\mathrm{WT}}\,.  
\end{eqnarray*}
\underline{Comparison with the model of Nacher, Oset, Ramos and Toki (NORT):}\\
\quad\\
In \cite{Nacher:1998mi}, the reaction $\gamma p\rightarrow K^{+}\pi^{0}\Sigma^{0}$ was studied, taking only the contact term from the WT class of graphs into account. In our present formalism, this leads to  $\mathcal{M}^{\mathrm{NORT}}_{1,\mathrm{tree}}=e/(8F_{\pi}F_{K})$, and, approaching the threshold,
\begin{equation}\label{eq:A10pNORT}
\mathcal{A}_{0+,\mathrm{NORT}}^{1(\mathrm{tree})} = \frac{e}{8F_{\pi}F_{K}}\sqrt{\frac{E_{\Sigma}^{\ast}+m_{\Sigma}}{2M_{\pi\Sigma}}}\sqrt{\frac{E_{\pi\Sigma}+M_{\pi\Sigma}}{2\sqrt{s}}}\left(\sqrt{s}-m_{N}\right) \,\rightarrow\, \frac{e(M_{\pi}+M_{K})}{8F_{\pi}F_{K}} + \mathcal{O}(M^2)\,.
\end{equation}
Comparing with our Eq.~(\ref{eq:A0pthrPi0S0}), we see that the inclusion of the rest of the WT class graphs would lead to the replacement $M_{K}\rightarrow M_{\pi}$ on the r.h.s of (\ref{eq:A10pNORT})\,.
In our formalism, it is impossible to have only $\mathcal{M}_{1}\not=0$ (as the contact term alone would be represented) because of the gauge invariance constraints. However, we can choose e.g. $\mathcal{M}_{8}=2\mathcal{M}_{1}/(t_{K}-M_{K}^2)$ and set all other $\mathcal{M}_{i\not=1,8}$ to zero. This will lead to an additional $\theta_{K}$-dependence of the resulting cross section, but it is suppressed at low energies. Up to the relativistic corrections we include, our formalism yields the same results for the case of the simple WT contact term photoproduction kernel as the one of \cite{Nacher:1998mi}\,, as we checked taking into account our different normalization of spinors and the correspondences
\begin{equation*}
-\sqrt{2m_{B'}}\frac{T^{\mathrm{NORT}}(M_{\pi\Sigma})}{8\pi M_{\pi\Sigma}}\sqrt{2m_{B}}\,\leftrightarrow\, f_{0+}(M_{\pi\Sigma})\,,\quad -\frac{G^{\mathrm{NORT}}(M_{\pi\Sigma})}{2m_{B}}\,\leftrightarrow\,G(M_{\pi\Sigma})\,,\quad C_{ij}\,\leftrightarrow\, (g_{WT})_{ij}\,.
\end{equation*}
\\
%

\section{Summary and outlook}
\label{sec:summary}

In this article, we have derived low-energy theorems for the two-meson photoproduction cross sections $\gamma p\rightarrow K^{+}MB$, $MB=\pi\Sigma,\,\bar{K}N,\,\pi\Lambda$, and have outlined a formalism to incorporate the meson-baryon final-state interaction through partial-wave amplitudes $f_{\ell\pm}$ constructed within the framework of ``unitarized ChPT''. This formalism can be used to implement coupled-channel unitarity, low-energy theorems from ChPT and gauge invariance in the description of the photoproduction process. It is now a natural question to ask for the other kinds of final-state interaction inherent in the process, as e.g. the pion-kaon interaction, or irreducible three-body interactions. Here it is important to note two things. First, effects due to an enhanced final-state interaction in the other channels have already been subtracted in the data with which we want to compare our theoretical description \cite{Moriya:2013eb}. And second, the effects due to this subtraction on the $\pi\Sigma$ invariant mass spectrum were reported to be moderate. Therefore, it seems that a direct application of the approach proposed here to this subtracted data is a reasonable strategy. This application will be the content of our subsequent studies based on the present work. In the future, one should also make efforts for a more complete description of the photoproduction process, along the lines explained so far. This would enable us to compare our predictions with a more complete subset of the provided data, as e.g. the kaon angular distributions, which are presumably sensitive to final state interaction in sectors not considered in this work. It is also conceivable to adopt strategies as detailed in \cite{Borasoy:2005zg,Borasoy:2007ku,Mai:2012wy} to implement unitarity together with gauge invariance. In the case of two-meson photoproduction, however, the development of such a model can be expected to be a very demanding and highly complex endeavor. \\

\subsection*{Acknowledgement}

I thank Petr Byd\v{z}ovsk\'y and Ale\v{s} Ciepl\'y for valuable comments on the manuscript. 
This work was supported by the Czech Science Foundation GACR grant 19-19640S.

\clearpage

\begin{appendix}

\section{Lorentz transformation and kinematic relations}
\label{app:kinematics}
\def\theequation{\Alph{section}.\arabic{equation}}
\setcounter{equation}{0}

Let us start in the c.m. frame where $\vec{k}$ points in the $z$-direction, $\vec{q}_{K}$ lies in the $x-z$-plane, and $\theta_{K}$ is the angle between $\vec{q}_{K}$ and $\vec{k}$. We have $\vec{p}_{N}+\vec{k}=\vec{0}=\vec{q}_{K}+(\vec{q}_{\pi}+\vec{p}_{\Sigma})$ (by definition of the frame). 
Now apply the following Lorentz transformation to the four-vectors in this frame:
\begin{displaymath}
\Lambda^{\mu}_{\,\,\nu}=\left(\begin{array}{cccc} 1 & 0 & 0 & 0 \\ 0 & \cos\alpha & 0 & \sin\alpha \\ 0 & 0 & 1 & 0 \\ 0 & -\sin\alpha & 0 & \cos\alpha \end{array}\right)\left(\begin{array}{cccc} \cosh\xi & 0 & 0 & -\sinh\xi \\ 0 & 1 & 0 & 0 \\ 0 & 0 & 1 & 0 \\ -\sinh\xi & 0 & 0 & \cosh\xi \end{array}\right)\left(\begin{array}{cccc} 1 & 0 & 0 & 0 \\ 0 & \cos\theta_{K} & 0 & -\sin\theta_{K} \\ 0 & 0 & 1 & 0 \\ 0 & \sin\theta_{K} & 0 & \cos\theta_{K} \end{array}\right)\,,
\end{displaymath}
\begin{equation*}
  \tanh\xi=-\frac{|\vec{q}_{K}|}{\sqrt{s}-E_{K}}\,,\quad \cos\alpha = \frac{|\vec{q}_{K}|+(\sqrt{s}-E_{K})\cos\theta_{K}}{\sqrt{s}-E_{K}+|\vec{q}_{K}|\cos\theta_{K}}\,.
\end{equation*}
Note that $\sqrt{s}-E_{K}=E_{\pi}+E_{\Sigma}$. First, we find that
\begin{equation}
\Lambda^{\mu}_{\,\,\nu}(q_{\pi}+p_{\Sigma})^{\nu} = \left(\sqrt{(\sqrt{s}-E_{K})^2-|\vec{q}_{K}|^2},0,0,0\right)=\left(M_{\pi\Sigma},0,0,0\right)\,,
\end{equation}
which means that the above transformation does the job of transforming to the $\ast$ frame. Next,
\begin{equation}
\Lambda^{\mu}_{\,\,\nu}k^{\nu} = \Lambda^{\mu}_{\,\,\nu}\left(E_{\gamma},0,0,E_{\gamma}\right) = \left(E_{\gamma}^{\,\ast},0,0,E_{\gamma}^{\,\ast}\right)\,,\quad E_{\gamma}^{\,\ast}=\frac{E_{\gamma}}{M_{\pi\Sigma}}\left(\sqrt{s}-E_{K}+|\vec{q}_{K}|\cos\theta_{K}\right)\,.
\end{equation}
We also find
\begin{equation}
\Lambda^{\mu}_{\,\,\nu}q_{K}^{\nu} = \left(E_{K}^{\,\ast},|\vec{q}_{K}^{\,\ast}|\sin\alpha,0,|\vec{q}_{K}^{\,\ast}|\cos\alpha\right)\,, \quad \Lambda^{\mu}_{\,\,\nu}p_{N}^{\nu} = \left(E_{N}^{\,\ast},\vec{q}_{K}^{\,\ast}-\vec{k}^{\ast}\right)\,.
\end{equation}
We have thus identified the angle $\theta_{K}^{\,\ast}=\alpha$ between $\vec{k}^{\ast}$ and  $\vec{q}_{K}^{\,\ast}$ in the $\ast$ frame:
\begin{equation}
\cos\theta_{K}^{\ast} = \frac{|\vec{q}_{K}|+(\sqrt{s}-E_{K})\cos\theta_{K}}{\sqrt{s}-E_{K}+|\vec{q}_{K}|\cos\theta_{K}}\,,\quad\mathrm{or}\quad \sin\theta_{K}^{\ast}=\frac{M_{\pi\Sigma}\sin\theta_{K}}{\sqrt{s}-E_{K}+|\vec{q}_{K}|\cos\theta_{K}}\,.
\end{equation}
The above transformation was chosen such that $\vec{k}^{\ast}$ again points in the $z$-direction (of the transformed frame), and $\vec{q}_{K}^{\,\ast}$ lies in the according $x-z$ plane.\\
For the initial state, the kinematics is very simple in the c.m. frame, where $\vec{p}_{N}+\vec{k}=\vec{0}$ $\Rightarrow$ $|\vec{p}_{N}|=|\vec{k}|=E_{\gamma}=(s-m_{N}^2)/(2\sqrt{s})=\sqrt{s}-E_{N}$. Moreover, from the Lorentz invariance of $t_{K}$,
\begin{equation}
\frac{E_{\gamma}^{\,\ast}}{E_{\gamma}} = \frac{E_{K}-|\vec{q}_{K}|\cos\theta_{K}}{E_{K}^{\,\ast}-|\vec{q}_{K}^{\,\ast}|\cos\theta_{K}^{\ast}}\,.
\end{equation}
Combining the results above, together with Eqs.~(\ref{eq:qK}), (\ref{eq:qKast}), we obtain the useful relations
\begin{eqnarray}
  2M_{\pi\Sigma}E_{\gamma}^{\,\ast} &=& 2E_{\gamma}\left(\sqrt{s}-E_{K}+|\vec{q}_{K}|\cos\theta_{K}\right) = s-m_{N}^2+t_{K}-M_{K}^2\,,\\
  2M_{\pi\Sigma}E_{N}^{\,\ast} &=& 2(\sqrt{s}-E_{\gamma})(\sqrt{s}-E_{K})-2E_{\gamma}|\vec{q}_{K}|\cos\theta_{K} = M_{\pi\Sigma}^2+m_{N}^2-t_{K}\,.
\end{eqnarray}

\newpage

\section{Chiral building blocks and notation}
\label{app:chptnom}
\def\theequation{\Alph{section}.\arabic{equation}}
\setcounter{equation}{0}

The Goldstone boson fields are collected in a matrix $U=\exp\left(i\sqrt{2}\mathbf{\Phi}/F_{0}\right)$, where $F_{0}$ is the meson decay constant in the (three-flavor) chiral limit of vanishing light-quark masses, $m_{u,d,s}\rightarrow 0$, $F_{0}\approx 80\,\mathrm{MeV}$ \cite{Aoki:2016frl}, and
\begin{equation}\label{eq:Phi}
\mathbf{\Phi} = \left(
\begin{array}{ccc}
\frac{1}{\sqrt{2}}\pi^{0}+\frac{1}{\sqrt{6}}\eta_{8} & \pi^{+} & K^{+} \\
\pi^{-} & -\frac{1}{\sqrt{2}}\pi^{0}+\frac{1}{\sqrt{6}}\eta_{8} & K^{0} \\
K^{-} & \bar K^{0} & -\frac{2}{\sqrt{6}}\eta_{8} \end{array}\right),
\end{equation}

We also define $u=\sqrt{U}$, and
\begin{eqnarray*}
\nabla_{\mu}U &=& \partial_{\mu}U-i(v_{\mu}+a_{\mu})U+iU(v_{\mu}-a_{\mu})\,,\quad u_{\mu}=iu^{\dagger}\left(\nabla_{\mu}U\right)u^{\dagger}\,,\\
\Gamma^{\mu} &=& \frac{1}{2}\left(u^{\dagger}[\partial^{\mu}-i(v^{\mu}+a^{\mu})]u + u[\partial^{\mu}-i(v^{\mu}-a^{\mu})]u^{\dagger}\right)\,, \\
\chi &=& 2B_{0}(\hat{s}+i\hat{p})\,,\quad \chi_{\pm} = \left(u^{\dagger}\chi u^{\dagger}\pm u\chi^{\dagger}u\right)\,.
\end{eqnarray*}
Here, $v,a,\hat{s},\hat{p}$ are vector, axial-vector, scalar and pseudoscalar source fields, and $B_{0}$ is a low-energy constant, related to the light-quark condensate in the chiral limit \cite{Gasser:1984gg,Aoki:2016frl}. The brackets $\langle\cdots\rangle$ denote the trace in flavor space.  In this work, we set $\hat{s}=\mathrm{diag}(\hat{m},\hat{m},m_{s}),\,\hat{p}=0$, $a^{\mu}=0$, and $v^{\mu}=-eQA^{\mu}$, where $A^{\mu}$ is the vector potential from the photon field, $Q$ is the quark charge matrix (i.e., $\mathrm{diag}(\frac{2}{3},-\frac{1}{3},-\frac{1}{3})$), and $\hat{m}=\frac{1}{2}(m_{u}+m_{d})$. 
The baryon fields are collected in
\begin{equation} \label{eq:B}
B=B^{a}\lambda^{a} = \left(
\begin{array}{ccc}
\frac{1}{\sqrt{2}}\Sigma^{0}+\frac{1}{\sqrt{6}}\Lambda & \Sigma^{+} & p \\
\Sigma^{-} & -\frac{1}{\sqrt{2}}\Sigma^{0}+\frac{1}{\sqrt{6}}\Lambda & n \\
\Xi^{-} & \Xi^{0} & -\frac{2}{\sqrt{6}}\Lambda \end{array}\right)\,,
\end{equation}
and the covariant derivative $D^{\mu}$ acts as $\lbrack D^{\mu},B\rbrack:=\partial^{\mu}B + \lbrack\Gamma^{\mu},B\rbrack$.

\newpage

\section{Results for the tree graphs}
\label{app:treegraphs}
\def\theequation{\Alph{section}.\arabic{equation}}
\setcounter{equation}{0}

\subsection{Graphs with a WT vertex}
Explicitly, the Feynman graphs with the $WT$ coupling contribute as follows:
\begin{eqnarray*}
  \mathcal{M}_{1} &=& \left(\frac{eg_{WT}^{\pi K}}{4F_{\pi}F_{K}}\right)\biggl(Q_{K}-Q_{\pi} + Q_{p} - Q_{\Sigma}\left(1+2\frac{t_{K}-M_{K}^2}{u_{\Sigma}-m_{\Sigma}^2}\right)\biggr)\,,\\
  \mathcal{M}_{2} &=& 2\left(\frac{eg_{WT}^{\pi K}}{4F_{\pi}F_{K}}\right)(m_{N}-m_{\Sigma})\biggl(\frac{Q_{p}}{s-m_{N}^2} + \frac{Q_{\pi}}{t_{\pi}-M_{\pi}^2}\biggr)\,,\\
  \mathcal{M}_{3} &=& 2\left(\frac{eg_{WT}^{\pi K}}{4F_{\pi}F_{K}}\right)(m_{N}-m_{\Sigma})\biggl(\frac{Q_{\Sigma}}{u_{\Sigma}-m_{\Sigma}^2} - \frac{Q_{\pi}}{t_{\pi}-M_{\pi}^2}\biggr)\,,\\
  \mathcal{M}_{4} &=& 2\left(\frac{eg_{WT}^{\pi K}}{4F_{\pi}F_{K}}\right)(m_{N}-m_{\Sigma})\biggl(\frac{Q_{K}}{t_{K}-M_{K}^2} - \frac{Q_{\pi}}{t_{\pi}-M_{\pi}^2}\biggr)\,,\\
  \mathcal{M}_{5} &=& \left(\frac{eg_{WT}^{\pi K}}{4F_{\pi}F_{K}}\right)(m_{N}-m_{\Sigma})\biggl(\frac{Q_{p}}{s-m_{N}^2} + \frac{Q_{\Sigma}}{u_{\Sigma}-m_{\Sigma}^2}\biggr)\,,\quad \mathcal{M}_{6} = \mathcal{M}_{7} = 0\,,\\
  \mathcal{M}_{8} &=& 4\left(\frac{eg_{WT}^{\pi K}}{4F_{\pi}F_{K}}\right)\biggl(\frac{Q_{K}}{t_{K}-M_{K}^2} -\frac{Q_{\Sigma}}{u_{\Sigma}-m_{\Sigma}^2}\biggr)\,,\quad \mathcal{M}_{9} = 0\,,\\
  \mathcal{M}_{10} &=& -4\left(\frac{eg_{WT}^{\pi K}}{4F_{\pi}F_{K}}\right)\biggl(\frac{Q_{p}}{s-m_{N}^2} + \frac{Q_{\pi}}{t_{\pi}-M_{\pi}^2}\biggr)\,,\\
  \mathcal{M}_{11} &=& -4\left(\frac{eg_{WT}^{\pi K}}{4F_{\pi}F_{K}}\right)\biggl(\frac{Q_{\Sigma}}{u_{\Sigma}-m_{\Sigma}^2} - \frac{Q_{\pi}}{t_{\pi}-M_{\pi}^2}\biggr)\,,\\
  \mathcal{M}_{12} &=& -4\left(\frac{eg_{WT}^{\pi K}}{4F_{\pi}F_{K}}\right)\biggl(\frac{Q_{K}}{t_{K}-M_{K}^2} - \frac{Q_{\pi}}{t_{\pi}-M_{\pi}^2}\biggr)\,,\\
  \mathcal{M}_{13} &=& -2\left(\frac{eg_{WT}^{\pi K}}{4F_{\pi}F_{K}}\right)\biggl(\frac{Q_{p}}{s-m_{N}^2} + \frac{Q_{\Sigma}}{u_{\Sigma}-m_{\Sigma}^2}\biggr)\,,\\
  \mathcal{M}_{14} &=& \mathcal{M}_{15} = \mathcal{M}_{16}= 0\,.
\end{eqnarray*}
We have used the abbreviation $t_{\pi}:=(q_{\pi}-k)^2$, and have replaced $F_{0}^2\rightarrow F_{\pi}F_{K}$. The $Q_{\cdots}$ denote the charge of the indicated particle in units of $e$, e.g. $Q_{p}=+1$. Using charge conservation,
\begin{displaymath}
  Q_{p}=Q_{\Sigma}+Q_{\pi}+Q_{K}\,,
\end{displaymath}
and the kinematic relation
\begin{displaymath}
  s-m_{N}^2 \,+\, u_{\Sigma}-m_{\Sigma}^2 \,+\, t_{K}-M_{K}^2 \,+\, t_{\pi}-M_{\pi}^2=0\,,
\end{displaymath}
 it is easy to verify that the constraints imposed by gauge invariance are fulfilled by the above set of amplitudes. The numerical values of the coupling strengths $g_{WT}$ are given at the end of the next subsection. We don't display the amplitudes $\mathcal{N}_{i}$ here, as they do not contribute to photoproduction observables.

 \newpage
 
\subsection{Born graphs}
\underline{Class B1: pion emitted first, followed by kaon.}
\begin{eqnarray*}
  \mathcal{M}_{1} &=& \left(\frac{eg^{K}_{\Sigma H}g^{\pi}_{HN}}{F_{\pi}F_{K}}\right)\left(Q_{K}-Q_{\Sigma}\frac{t_{K}-M_{K}^2}{u_{\Sigma}-m_{\Sigma}^2}\right)\left(1+\frac{(m_{N}+m_{H})(m_{\Sigma}+m_{H})}{(p_{N}-q_{\pi})^2-m_{H}^2}\right)\,,\\
  \mathcal{M}_{2} &=& 2(m_{H}+m_{N})\left(\frac{eg^{K}_{\Sigma H}g^{\pi}_{HN}}{F_{\pi}F_{K}}\right)\left(\frac{M_{K\Sigma}^2-m_{\Sigma}^2}{M_{K\Sigma}^2-m_{H}^2}\right)\left(\frac{Q_{p}}{s-m_{N}^2} + \frac{Q_{\pi}}{t_{\pi}-M_{\pi}^2}\right)\,,\\
  \mathcal{M}_{3} &=& \left(\frac{eg^{K}_{\Sigma H}g^{\pi}_{HN}}{F_{\pi}F_{K}}\right)\left(\frac{2(m_{H}+m_{N})}{u_{\Sigma}-m_{\Sigma}^2}\right)\biggl(Q_{\Sigma}\left(\frac{M_{K\Sigma}^2-m_{\Sigma}^2}{M_{K\Sigma}^2-m_{H}^2}+\frac{(t_{K}-M_{K}^2)(m_{\Sigma}^2-m_{H}^2)}{(M_{K\Sigma}^2-m_{H}^2)((p_{N}-q_{\pi})^2-m_{H}^2)}\right)\\
  &\quad& \hspace{4.6cm} -\,\,Q_{K}\biggl(\frac{(u_{\Sigma}-m_{\Sigma}^2)(m_{\Sigma}^2-m_{H}^2)}{(M_{K\Sigma}^2-m_{H}^2)((p_{N}-q_{\pi})^2-m_{H}^2)}\biggr)\\
  &\quad& \hspace{4.6cm} -\,\,Q_{\pi}\left(\frac{M_{K\Sigma}^2-m_{\Sigma}^2}{M_{K\Sigma}^2-m_{H}^2}\right)\left(\frac{u_{\Sigma}-m_{\Sigma}^2}{t_{\pi}-M_{\pi}^2}\right)\biggr)\,,\\
  \mathcal{M}_{4} &=& \left(\frac{eg^{K}_{\Sigma H}g^{\pi}_{HN}}{F_{\pi}F_{K}}\right)\left(\frac{2(m_{H}+m_{N})}{t_{K}-M_{K}^2}\right)\biggl(Q_{K}\biggl(1+\left(\frac{m_{\Sigma}^2-m_{H}^2}{M_{K\Sigma}^2-m_{H}^2}\right)\left(\frac{m_{H}^2+M_{K}^2-t_{K}-M_{K\Sigma}^2}{(p_{N}-q_{\pi})^2-m_{H}^2}\right)\biggr)\\
  &\quad& \hspace{4.6cm} -\,\,Q_{\Sigma}\left(\frac{m_{\Sigma}^2-m_{H}^2}{M_{K\Sigma}^2-m_{H}^2}\right)\biggl(\frac{t_{K}-M_{K}^2}{(p_{N}-q_{\pi})^2-m_{H}^2}\biggr)\\
  &\quad& \hspace{4.6cm} -\,\,Q_{\pi}\left(\frac{M_{K\Sigma}^2-m_{\Sigma}^2}{M_{K\Sigma}^2-m_{H}^2}\right)\left(\frac{t_{K}-M_{K}^2}{t_{\pi}-M_{\pi}^2}\right)\biggr)\,,\\
  \mathcal{M}_{5} &=& (m_{H}+m_{N})\left(\frac{eg^{K}_{\Sigma H}g^{\pi}_{HN}}{F_{\pi}F_{K}}\right)\biggl(\frac{Q_{p}}{s-m_{N}^2}\left(\frac{M_{K\Sigma}^2-m_{\Sigma}^2}{M_{K\Sigma}^2-m_{H}^2}\right) + \frac{Q_{\Sigma}}{u_{\Sigma}-m_{\Sigma}^2}\left(\frac{(p_{N}-q_{\pi})^2-m_{\Sigma}^2}{(p_{N}-q_{\pi})^2-m_{H}^2}\right)\\
  &\quad& \hspace{4.6cm} -\,\,\frac{(Q_{K}+Q_{\Sigma})(m_{\Sigma}^2-m_{H}^2)}{((p_{N}-q_{\pi})^2-m_{H}^2)(M_{K\Sigma}^2-m_{H}^2)}\biggr)\,,\\
  \mathcal{M}_{6} &=& \mathcal{M}_{7} = 0\,,\qquad \mathcal{M}_{8} = \frac{2}{t_{K}-M_{K}^2}\mathcal{M}_{1}\,,\qquad \mathcal{M}_{9}=0\,,\\
  \mathcal{M}_{10} &=& -2\left(\frac{eg^{K}_{\Sigma H}g^{\pi}_{HN}}{F_{\pi}F_{K}}\right)\left(\frac{M_{K\Sigma}^2-m_{\Sigma}^2+(m_{\Sigma}+m_{H})(m_{\Sigma}+m_{N})}{M_{K\Sigma}^2-m_{H}^2}\right)\biggl(\frac{Q_{p}}{s-m_{N}^2}+\frac{Q_{\pi}}{t_{\pi}-M_{\pi}^2}\biggr)\,,\\
  \mathcal{M}_{11} &=& -2\left(\frac{eg^{K}_{\Sigma H}g^{\pi}_{HN}}{F_{\pi}F_{K}}\right)\biggl(\frac{Q_{\Sigma}}{u_{\Sigma}-m_{\Sigma}^2} + \frac{(m_{N}+m_{H})(m_{\Sigma}+m_{H})}{(p_{N}-q_{\pi})^2-m_{H}^2}\left(\frac{Q_{\Sigma}}{u_{\Sigma}-m_{\Sigma}^2} + \frac{Q_{K}+Q_{\Sigma}}{M_{K\Sigma}^2-m_{H}^2}\right)\\
  &\quad& \hspace{4.6cm} -\,\,\left(\frac{M_{K\Sigma}^2-m_{\Sigma}^2+(m_{\Sigma}+m_{H})(m_{\Sigma}+m_{N})}{M_{K\Sigma}^2-m_{H}^2}\right)\left(\frac{Q_{\pi}}{t_{\pi}-M_{\pi}^2}\right)\biggr)\,,\\
  \mathcal{M}_{12} &=& -2\left(\frac{eg^{K}_{\Sigma H}g^{\pi}_{HN}}{F_{\pi}F_{K}}\right)\biggl(\frac{Q_{K}}{t_{K}-M_{K}^2} + \frac{(m_{N}+m_{H})(m_{\Sigma}+m_{H})}{(p_{N}-q_{\pi})^2-m_{H}^2}\left(\frac{Q_{K}}{t_{K}-M_{K}^2} + \frac{Q_{K}+Q_{\Sigma}}{M_{K\Sigma}^2-m_{H}^2}\right)\\
    &\quad& \hspace{4.6cm} -\,\,\left(\frac{M_{K\Sigma}^2-m_{\Sigma}^2+(m_{\Sigma}+m_{H})(m_{\Sigma}+m_{N})}{M_{K\Sigma}^2-m_{H}^2}\right)\left(\frac{Q_{\pi}}{t_{\pi}-M_{\pi}^2}\right)\biggr)\,,\\
  \mathcal{M}_{13} &=& -\left(\frac{eg^{K}_{\Sigma H}g^{\pi}_{HN}}{F_{\pi}F_{K}}\right)\biggl(\left(\frac{M_{K\Sigma}^2-m_{\Sigma}^2+(m_{\Sigma}+m_{H})(m_{\Sigma}+m_{N})}{M_{K\Sigma}^2-m_{H}^2}\right)\left(\frac{Q_{p}}{s-m_{N}^2}\right) + \frac{Q_{\Sigma}}{u_{\Sigma}-m_{\Sigma}^2}\\
  &\quad& \hspace{4.6cm} +\,\,\frac{(m_{N}+m_{H})(m_{\Sigma}+m_{H})}{(p_{N}-q_{\pi})^2-m_{H}^2}\left(\frac{Q_{\Sigma}}{u_{\Sigma}-m_{\Sigma}^2} + \frac{Q_{K}+Q_{\Sigma}}{M_{K\Sigma}^2-m_{H}^2}\right)\biggr)\,,
 \end{eqnarray*}
and $\mathcal{M}_{14} = \mathcal{M}_{15} = \mathcal{M}_{16} = 0\,$. \\
\quad \\
\underline{Class B2: kaon emitted first, followed by pion.}
\begin{eqnarray*}
  \mathcal{M}_{1} &=& \left(\frac{eg^{\pi}_{\Sigma H}g^{K}_{HN}}{F_{\pi}F_{K}}\right)\biggl(\left(\frac{t_{K}-M_{K}^2}{(p_{N}-q_{K})^2-m_{H}^2}\right)\biggl(\frac{(Q_{\pi}+Q_{\Sigma})(m_{\Sigma}+m_{H})(m_{N}+m_{H})}{M_{\pi\Sigma}^2-m_{H}^2} \\  &\quad&  \hspace{6.2cm} +\,\,\frac{Q_{\Sigma}((p_{N}-q_{K})^2-m_{N}^2+(m_{H}+m_{N})(m_{\Sigma}+m_{N}))}{u_{\Sigma}-m_{\Sigma}^2}\biggr) \\
  &\quad&  \hspace{6.2cm} -\,\,\frac{Q_{K}(M_{\pi\Sigma}^2-m_{\Sigma}^2+(m_{\Sigma}+m_{H})(m_{\Sigma}+m_{N}))}{M_{\pi\Sigma}^2-m_{H}^2}\biggr)\,,\\
  \mathcal{M}_{2} &=& 2(m_{\Sigma}+m_{H})\left(\frac{eg^{\pi}_{\Sigma H}g^{K}_{HN}}{F_{\pi}F_{K}}\right)\biggl(\frac{Q_{p}}{s-m_{N}^2}\left(\frac{M_{\pi\Sigma}^2-s}{M_{\pi\Sigma}^2-m_{H}^2}\right) + \frac{Q_{K}}{M_{\pi\Sigma}^2-m_{H}^2} \\  &\quad&  \hspace{4.6cm} +\,\,\left(\frac{(p_{N}-q_{K})^2-m_{N}^2}{(p_{N}-q_{K})^2-m_{H}^2}\right)\left(\frac{Q_{\pi}+Q_{\Sigma}}{M_{\pi\Sigma}^2-m_{H}^2}+\frac{Q_{\pi}}{t_{\pi}-M_{\pi}^2}\right)\biggr)\,,\\
  \mathcal{M}_{3} &=& 2(m_{\Sigma}+m_{H})\left(\frac{eg^{\pi}_{\Sigma H}g^{K}_{HN}}{F_{\pi}F_{K}}\right)\left(\frac{(p_{N}-q_{K})^2-m_{N}^2}{(p_{N}-q_{K})^2-m_{H}^2}\right)\left(\frac{Q_{\Sigma}}{u_{\Sigma}-m_{\Sigma}^2}-\frac{Q_{\pi}}{t_{\pi}-M_{\pi}^2}\right)\,,\\
  \mathcal{M}_{4} &=& 2(m_{\Sigma}+m_{H})\left(\frac{eg^{\pi}_{\Sigma H}g^{K}_{HN}}{F_{\pi}F_{K}}\right)\biggl(\frac{Q_{K}}{t_{K}-M_{K}^2}\left(\frac{M_{\pi\Sigma}^2-m_{N}^2}{M_{\pi\Sigma}^2-m_{H}^2}\right) + \frac{Q_{\pi}+Q_{\Sigma}}{M_{\pi\Sigma}^2-m_{H}^2}\left(\frac{m_{N}^2-m_{H}^2}{(p_{N}-q_{K})^2-m_{H}^2}\right)\\  &\quad&  \hspace{4.6cm} -\,\,\frac{Q_{\pi}}{t_{\pi}-M_{\pi}^2}\left(\frac{(p_{N}-q_{K})^2-m_{N}^2}{(p_{N}-q_{K})^2-m_{H}^2}\right)\biggr)\,,\\
  \mathcal{M}_{5} &=& (m_{\Sigma}+m_{H})\left(\frac{eg^{\pi}_{\Sigma H}g^{K}_{HN}}{F_{\pi}F_{K}}\right)\biggl(\frac{Q_{p}}{s-m_{N}^2}\left(\frac{M_{\pi\Sigma}^2-s}{M_{\pi\Sigma}^2-m_{H}^2}\right) + \frac{Q_{K}}{M_{\pi\Sigma}^2-m_{H}^2} \\  &\quad&  \hspace{4.6cm} +\,\,\left(\frac{(p_{N}-q_{K})^2-m_{N}^2}{(p_{N}-q_{K})^2-m_{H}^2}\right)\left(\frac{Q_{\pi}+Q_{\Sigma}}{M_{\pi\Sigma}^2-m_{H}^2}+\frac{Q_{\Sigma}}{u_{\Sigma}-m_{\Sigma}^2}\right)\biggr)\,,\\
  \mathcal{M}_{6} &=& \mathcal{M}_{7} = 0\,,\qquad \mathcal{M}_{8} = \frac{2}{t_{K}-M_{K}^2}\mathcal{M}_{1}\,,\qquad \mathcal{M}_{9}=0\,,\\
 \mathcal{M}_{10} &=& \left(\frac{eg^{\pi}_{\Sigma H}g^{K}_{HN}}{F_{\pi}F_{K}}\right)\biggl(\left(\frac{2}{(p_{N}-q_{K})^2-m_{H}^2}\right)\biggl(\frac{(Q_{\pi}+Q_{\Sigma})(m_{\Sigma}+m_{H})(m_{N}+m_{H})}{M_{\pi\Sigma}^2-m_{H}^2} \\  &\quad&  \hspace{6.2cm} +\,\,\frac{Q_{\pi}((p_{N}-q_{K})^2-m_{N}^2+(m_{H}+m_{N})(m_{\Sigma}+m_{N}))}{t_{\pi}-M_{\pi}^2}\biggr) \\
 &\quad&  \hspace{6.2cm} +\,\,\frac{2Q_{p}(M_{\pi\Sigma}^2-m_{\Sigma}^2+(m_{\Sigma}+m_{H})(m_{\Sigma}+m_{N}))}{(s-m_{N}^2)(M_{\pi\Sigma}^2-m_{H}^2)}\biggr)\,,\\
 \mathcal{M}_{11} &=& 2\left(\frac{eg^{\pi}_{\Sigma H}g^{K}_{HN}}{F_{\pi}F_{K}}\right)\left(\frac{(p_{N}-q_{K})^2-m_{N}^2+(m_{H}+m_{N})(m_{\Sigma}+m_{N})}{(p_{N}-q_{K})^2-m_{H}^2}\right)\left(\frac{Q_{\Sigma}}{u_{\Sigma}-m_{\Sigma}^2}-\frac{Q_{\pi}}{t_{\pi}-M_{\pi}^2}\right)\,,\\
 \mathcal{M}_{12} &=& -\left(\frac{eg^{\pi}_{\Sigma H}g^{K}_{HN}}{F_{\pi}F_{K}}\right)\biggl(\left(\frac{2}{(p_{N}-q_{K})^2-m_{H}^2}\right)\biggl(\frac{(Q_{\pi}+Q_{\Sigma})(m_{\Sigma}+m_{H})(m_{N}+m_{H})}{M_{\pi\Sigma}^2-m_{H}^2} \\  &\quad&  \hspace{6.2cm} +\,\,\frac{Q_{\pi}((p_{N}-q_{K})^2-m_{N}^2+(m_{H}+m_{N})(m_{\Sigma}+m_{N}))}{t_{\pi}-M_{\pi}^2}\biggr) \\
 &\quad&  \hspace{6.2cm} -\,\,\frac{2Q_{K}(M_{\pi\Sigma}^2-m_{\Sigma}^2+(m_{\Sigma}+m_{H})(m_{\Sigma}+m_{N}))}{(t_{K}-M_{K}^2)(M_{\pi\Sigma}^2-m_{H}^2)}\biggr)\,,\\
 \mathcal{M}_{13} &=& \left(\frac{eg^{\pi}_{\Sigma H}g^{K}_{HN}}{F_{\pi}F_{K}}\right)\biggl(\left(\frac{1}{(p_{N}-q_{K})^2-m_{H}^2}\right)\biggl(\frac{(Q_{\pi}+Q_{\Sigma})(m_{\Sigma}+m_{H})(m_{N}+m_{H})}{M_{\pi\Sigma}^2-m_{H}^2} \\  &\quad&  \hspace{6.2cm} +\,\,\frac{Q_{\Sigma}((p_{N}-q_{K})^2-m_{N}^2+(m_{H}+m_{N})(m_{\Sigma}+m_{N}))}{u_{\Sigma}-m_{\Sigma}^2}\biggr) \\
  &\quad&  \hspace{6.2cm} +\,\,\frac{Q_{p}(M_{\pi\Sigma}^2-m_{\Sigma}^2+(m_{\Sigma}+m_{H})(m_{\Sigma}+m_{N}))}{(s-m_{N}^2)(M_{\pi\Sigma}^2-m_{H}^2)}\biggr)\,,\\
\end{eqnarray*}
and, again, $\mathcal{M}_{14} = \mathcal{M}_{15} = \mathcal{M}_{16} = 0\,$. $H$ labels the intermediate baryon from $p\rightarrow\pi H$ or $p\rightarrow KH$ (all graphs have to be summed over the possible intermediate one-baryon states $H$, e.g., for $\gamma p\rightarrow K^{+}\pi\Sigma$, $H$ runs over $\Sigma^{0},\Lambda$ for the B2 graphs). 
 - One can use the following kinematic relations,
\begin{eqnarray*}
  (k-q_{\pi})^2 &=:& t_{\pi}\,,\quad s-m_{N}^2 + u_{\Sigma}-m_{\Sigma}^2 + t_{K}-M_{K}^2 + t_{\pi}-M_{\pi}^2=0\,,\\
  (p_{N}-q_{\pi})^2  &=& M_{K\Sigma}^2 + u_{\Sigma}-m_{\Sigma}^2 + t_{K}-M_{K}^2\,,\\
  (p_{N}-q_{K})^2 &=& M_{\pi\Sigma}^2 + m_{N}^2 -s + M_{K}^2-t_{K}\,,
\end{eqnarray*}
to verify the gauge invariance constraints for the two classes of Born graphs.\\ 
At the overall threshold $s=s_{thr}:=(m_{\Sigma}+M_{\pi}+M_{K})^2$, $M_{\pi\Sigma}=m_{\Sigma}+M_{\pi}$, we find
\begin{eqnarray*}
  M_{K\Sigma}^2 &\rightarrow& (m_{\Sigma}+M_{K})^2\,,\quad u_{\Sigma}-m_{\Sigma}^2 \,\rightarrow\, -\frac{m_{\Sigma}}{\sqrt{s_{thr}}}(s_{thr}-m_{N}^2)\,,\\
  t_{K}-M_{K}^2 &\rightarrow& -\frac{M_{K}}{\sqrt{s_{thr}}}(s_{thr}-m_{N}^2)\,,\quad t_{\pi}-M_{\pi}^2\,\rightarrow\, -\frac{M_{\pi}}{\sqrt{s_{thr}}}(s_{thr}-m_{N}^2)\,,\\
  t_{\Sigma}-m_{\Sigma}^2 &\rightarrow& m_{N}^2-\frac{m_{\Sigma}}{\sqrt{s_{thr}}}(s_{thr}+m_{N}^2)\,,\\
  (p_{N}-q_{\pi})^2 &\rightarrow& \frac{m_{\Sigma}+M_{K}}{\sqrt{s_{thr}}}\left(m_{N}^2 - M_{\pi}(m_{\Sigma}+M_{\pi}+M_{K})\right)\,,\\
  (p_{N}-q_{K})^2 &\rightarrow& \frac{m_{\Sigma}+M_{\pi}}{\sqrt{s_{thr}}}\left(m_{N}^2 - M_{K}(m_{\Sigma}+M_{\pi}+M_{K})\right)\,.
\end{eqnarray*}
For other meson-baryon pairs $MB$ than $\pi\Sigma$, one can simply replace $M_{\pi}$, $m_{\Sigma}$ by the pertaining masses, adjust the coupling factors given below, as well as the charges $Q$ and the meson decay constants, and sum over $H$ accordingly. Explicitly, we find from the vertex rules of the Lagrangian in Eq.~(\ref{eq:MB_LagrBMW}):
\begin{eqnarray*}
  g^{\pi^{0}}_{pp} &=& \frac{D+F}{2}\,,\quad g^{\pi^{+}}_{np} = \frac{D+F}{\sqrt{2}}\,,\quad g^{K^{+}}_{\Sigma^{0}p} = \frac{D-F}{2}\,,\quad g^{K^{+}}_{\Sigma^{-}n} = \frac{D-F}{\sqrt{2}}\,,\quad g^{K^{+}}_{\Lambda p} = -\frac{D+3F}{2\sqrt{3}}\,,\\
  g^{\pi^{0}}_{\Sigma^{0}\Sigma^{0}} &=& 0\,,\quad g^{\pi^{+}}_{\Sigma^{-}\Sigma^{0}} = F\,,\quad g^{\pi^{-}}_{\Sigma^{+}\Sigma^{0}} = -F\,,\quad g^{\pi^{0}}_{\Sigma^{0}\Lambda} = g^{\pi^{+}}_{\Sigma^{-}\Lambda} = g^{\pi^{-}}_{\Sigma^{+}\Lambda} = \frac{D}{\sqrt{3}}\,,\quad g^{\pi^{0}}_{\Lambda\Lambda} = 0\,,\\
  g^{K^{-}}_{p\Sigma^{0}} &=& \frac{D-F}{2}\,,\quad g^{\bar{K}^{0}}_{n\Sigma^{0}} = \frac{F-D}{2}\,,\quad g^{K^{-}}_{p\Lambda} = g^{\bar{K}^{0}}_{n\Lambda} = -\frac{D+3F}{2\sqrt{3}}\,,
\end{eqnarray*}
and for the Weinberg-Tomozawa couplings:
\begin{eqnarray*}
  g_{WT}^{\pi K}(\gamma p\rightarrow K^{+}\pi^{0}\Sigma^{0}) &=& \frac{1}{2}\,,\quad g_{WT}^{\pi K}(\gamma p\rightarrow K^{+}\pi^{+}\Sigma^{-}) = 0\,,\quad g_{WT}^{\pi K}(\gamma p\rightarrow K^{+}\pi^{-}\Sigma^{+}) = 1\,,\\
  g_{WT}^{\pi K}(\gamma p\rightarrow K^{+}\pi^{0}\Lambda) &=& \frac{\sqrt{3}}{2}\,,\quad g_{WT}^{\bar{K}K}(\gamma p\rightarrow K^{+}K^{-}p) = 2\,,\quad \,g_{WT}^{\bar{K}K}(\gamma p\rightarrow K^{+}\bar{K}^{0}n) = 1\,.
\end{eqnarray*}
\newpage

\subsection{Contribution from the anomalous Lagrangian}

For the process $\gamma K^{+}(l)\rightarrow K^{+}(q_{K})\pi^{0}(q_{\pi})$ (of odd intrinsic parity), one obtains a vertex rule from the anomalous Wess-Zumino-Witten-Lagrangian,
\begin{displaymath}
\frac{ieN_{c}}{12\pi^2F_{0}^3}\epsilon^{\mu\nu\rho\sigma}l_{\nu}(q_{K})_{\rho}(q_{\pi})_{\sigma}\,,
\end{displaymath}
where $N_{c}$ is the number of colours in QCD (there is a zero result for $\gamma K^{0}\rightarrow K^{+}\pi^{-}$). This vertex satisfies gauge invariance even off the mass shell because of the presence of the Levi-Civita tensor and $k=q_{K}+q_{\pi}-l\,$. For the graph contributing to $\gamma p\rightarrow \pi^{0}\Sigma^{0}K^{+}$, we find\footnote{For a $\Lambda$ instead of $\Sigma^{0}$, one has to replace $(D-F)\rightarrow -(D+3F)/\sqrt{3}\,$, $m_{\Sigma}\rightarrow m_{\Lambda}$.}
\begin{displaymath}
\frac{eN_{c}(D-F)}{24\pi^2F_{0}^4}\frac{l_{\nu}(q_{K})_{\rho}(q_{\pi})_{\sigma}}{l^2-M_{K}^2}(\slashed{p}_{\Sigma}-\slashed{p}_{N})\epsilon^{\mu\nu\rho\sigma}\gamma_{5}\,, \qquad l=p_{N}-p_{\Sigma}\,.
\end{displaymath}
Now we can go to the mass shell and use the identity
\begin{displaymath}
\epsilon^{\mu\nu\rho\sigma}\gamma_{5}=\frac{i}{4!}\left(\gamma^{\mu}\gamma^{\nu}\gamma^{\rho}\gamma^{\sigma}\mp (\mathrm{antisymm.\,\,perm.})\right) \equiv i\gamma^{\mu\nu\rho\sigma}\,,
\end{displaymath}
($\epsilon^{0123}=1$, $\gamma_{5}=i\gamma^{0}\gamma^{1}\gamma^{2}\gamma^{3}$) to obtain the resulting amplitudes $\mathcal{M}_{i}$:
\begin{eqnarray*}
  \mathcal{M}_{1}^{\mathrm{ano}} &=& \frac{g_{\mathrm{ano}}}{M_{K}^2-t_{\Sigma}}(m_{N}+m_{\Sigma})(M_{K}^2-t_{K})\,,\quad   \mathcal{M}_{2}^{\mathrm{ano}} =  \mathcal{M}_{3}^{\mathrm{ano}} = \frac{g_{\mathrm{ano}}}{M_{K}^2-t_{\Sigma}}(t_{K}-M_{K}^2)\,,\\
  \mathcal{M}_{4}^{\mathrm{ano}} &=& \frac{g_{\mathrm{ano}}}{M_{K}^2-t_{\Sigma}}(s-m_{N}^2+m_{\Sigma}^2-u_{\Sigma})\,,\\
  \mathcal{M}_{5}^{\mathrm{ano}} &=& \frac{g_{\mathrm{ano}}}{M_{K}^2-t_{\Sigma}}(s-M_{\pi\Sigma}^2+t_{K}-M_{K}^2+M_{K\Sigma}^2-m_{\Sigma}^2)\,,\quad \mathcal{M}_{6}^{\mathrm{ano}} = \mathcal{M}_{7}^{\mathrm{ano}} = 0\,,\\
  \mathcal{M}_{8}^{\mathrm{ano}} &=& \frac{2\mathcal{M}_{1}^{\mathrm{ano}}}{t_{K}-M_{K}^2}\,,\quad \mathcal{M}_{9}^{\mathrm{ano}} = -\mathcal{M}_{4}^{\mathrm{ano}}\,,\quad \mathcal{M}_{10}^{\mathrm{ano}} = \mathcal{M}_{11}^{\mathrm{ano}}  = \mathcal{M}_{12}^{\mathrm{ano}} = 0\,,\\
   \mathcal{M}_{13}^{\mathrm{ano}} &=& -\frac{2g_{\mathrm{ano}}}{M_{K}^2-t_{\Sigma}}(m_{N}+m_{\Sigma})\,,\quad \mathcal{M}_{14}^{\mathrm{ano}} = \mathcal{M}_{15}^{\mathrm{ano}} = \frac{2g_{\mathrm{ano}}}{M_{K}^2-t_{\Sigma}}\,,\quad \mathcal{M}_{16}^{\mathrm{ano}} = 0\,,
\end{eqnarray*}
where $g_{\mathrm{ano}}=eN_{c}(D-F)(m_{N}+m_{\Sigma})/(48\pi^2F_{\pi}F_{K}^3)$\,, again replacing $F_{0}$ by the appropriate physical decay constants. This graph does not contribute at the reaction threshold.

\newpage

\section{Projection on $\mathbf{\pi\Sigma}$ partial waves}
\label{app:cc_formalism}
\def\theequation{\Alph{section}.\arabic{equation}}
\setcounter{equation}{0}

We give the explicit expressions for the partial-wave projections of the photoproduction amplitude for $MB=\pi\Sigma$, but it is obvious that the analogous expressions for the other meson-baryon-channels can be found by simple replacements of the appropriate masses. As a first step, we define the combinations
\begin{eqnarray*}
  \mathcal{C}_{0+}^{1} &=& \mathcal{M}_{1}' + (\sqrt{s}+m_{N})\mathcal{M}_{5}'+(\sqrt{s}-M_{\pi\Sigma})\left(\mathcal{M}_{9}'+(\sqrt{s}+m_{N})\mathcal{M}_{13}'\right)\,,\\
  \mathcal{C}_{0+}^{2} &=& \mathcal{M}_{1}' - (\sqrt{s}-m_{N})\mathcal{M}_{5}' - (\sqrt{s}+M_{\pi\Sigma})\left(\mathcal{M}_{9}'-(\sqrt{s}-m_{N})\mathcal{M}_{13}'\right)\,,\\
  \mathcal{C}_{0+}^{3} &=& \mathcal{M}_{1}' + (\sqrt{s}-M_{\pi\Sigma})\mathcal{M}_{9}' \\ &+& \frac{1}{2}(\sqrt{s}+m_{N})\left(\mathcal{M}_{2}'+(\sqrt{s}-m_{N})\mathcal{M}_{6}'+(\sqrt{s}-M_{\pi\Sigma})\left(\mathcal{M}_{10}'+(\sqrt{s}-m_{N})\mathcal{M}_{14}'\right)\right)\,,\\
  \mathcal{C}_{0+}^{4} &=& \mathcal{M}_{1}' - (\sqrt{s}+M_{\pi\Sigma})\mathcal{M}_{9}' \\ &-& \frac{1}{2}(\sqrt{s}-m_{N})\left(\mathcal{M}_{2}'-(\sqrt{s}+m_{N})\mathcal{M}_{6}'-(\sqrt{s}+M_{\pi\Sigma})\left(\mathcal{M}_{10}'-(\sqrt{s}+m_{N})\mathcal{M}_{14}'\right)\right)\,,\\
  \mathcal{C}_{1-}^{1} &=& \mathcal{M}_{1}''-(\sqrt{s}-m_{N})\mathcal{M}_{5}'' - (\sqrt{s}-M_{\pi\Sigma})\left(\mathcal{M}_{9}''-(\sqrt{s}-m_{N})\mathcal{M}_{13}''\right)\,,\\
  \mathcal{C}_{1-}^{2} &=& \mathcal{M}_{1}'' + (\sqrt{s}+m_{N})\mathcal{M}_{5}''+(\sqrt{s}+M_{\pi\Sigma})\left(\mathcal{M}_{9}''+(\sqrt{s}+m_{N})\mathcal{M}_{13}''\right)\,,\\
  \mathcal{C}_{1-}^{3} &=& \mathcal{M}_{1}'' - (\sqrt{s}-M_{\pi\Sigma})\mathcal{M}_{9}'' \\ &-& \frac{1}{2}(\sqrt{s}-m_{N})\left(\mathcal{M}_{2}''-(\sqrt{s}+m_{N})\mathcal{M}_{6}''-(\sqrt{s}-M_{\pi\Sigma})\left(\mathcal{M}_{10}''-(\sqrt{s}+m_{N})\mathcal{M}_{14}''\right)\right)\,,\\
  \mathcal{C}_{1-}^{4} &=& \mathcal{M}_{1}'' + (\sqrt{s}+M_{\pi\Sigma})\mathcal{M}_{9}'' \\ &+& \frac{1}{2}(\sqrt{s}+m_{N})\left(\mathcal{M}_{2}''+(\sqrt{s}-m_{N})\mathcal{M}_{6}''+(\sqrt{s}+M_{\pi\Sigma})\left(\mathcal{M}_{10}''+(\sqrt{s}-m_{N})\mathcal{M}_{14}''\right)\right)\,,
\end{eqnarray*}
employing the following abbreviations:
\begin{eqnarray*}
  \mathcal{M}_{1}' &:=&  \overline{\mathcal{M}}_{1} - \frac{1}{3}(E_{\Sigma}^{\ast}-m_{\Sigma})\overline{\mathcal{M}}_{3}\,,\qquad \mathcal{M}_{1}'' :=  \overline{\mathcal{M}}_{1} + \frac{1}{3}(E_{\Sigma}^{\ast}+m_{\Sigma})\overline{\mathcal{M}}_{3}\,,\\
  \mathcal{M}_{5}' &:=&  \overline{\mathcal{M}}_{5} + \frac{1}{3}(E_{\Sigma}^{\ast}-m_{\Sigma})\overline{\mathcal{M}}_{7}\,,\qquad \mathcal{M}_{5}'' :=  \overline{\mathcal{M}}_{5} - \frac{1}{3}(E_{\Sigma}^{\ast}+m_{\Sigma})\overline{\mathcal{M}}_{7}\,,\\
  \mathcal{M}_{9}' &:=&  \overline{\mathcal{M}}_{9} + \frac{1}{3}(E_{\Sigma}^{\ast}-m_{\Sigma})\overline{\mathcal{M}}_{11}\,,\qquad \mathcal{M}_{9}'' :=  \overline{\mathcal{M}}_{9} - \frac{1}{3}(E_{\Sigma}^{\ast}+m_{\Sigma})\overline{\mathcal{M}}_{11}\,,\\
  \mathcal{M}_{13}' &:=&  \overline{\mathcal{M}}_{13} - \frac{1}{3}(E_{\Sigma}^{\ast}-m_{\Sigma})\overline{\mathcal{M}}_{15}\,,\qquad \mathcal{M}_{13}'' :=  \overline{\mathcal{M}}_{13} + \frac{1}{3}(E_{\Sigma}^{\ast}+m_{\Sigma})\overline{\mathcal{M}}_{15}\,,\\
  \mathcal{M}_{2}' &:=&  \overline{\mathcal{M}}_{2} + \frac{1}{3M_{\pi\Sigma}}(4E_{\Sigma}^{\ast}-m_{\Sigma})\overline{\mathcal{M}}_{3}\,,\qquad \mathcal{M}_{2}'' :=  \overline{\mathcal{M}}_{2} + \frac{1}{3M_{\pi\Sigma}}(4E_{\Sigma}^{\ast}+m_{\Sigma})\overline{\mathcal{M}}_{3}\,,\\
  \mathcal{M}_{6}' &:=&  \overline{\mathcal{M}}_{6} + \frac{1}{3M_{\pi\Sigma}}(4E_{\Sigma}^{\ast}-m_{\Sigma})\overline{\mathcal{M}}_{7}\,,\qquad \mathcal{M}_{6}'' :=  \overline{\mathcal{M}}_{6} + \frac{1}{3M_{\pi\Sigma}}(4E_{\Sigma}^{\ast}+m_{\Sigma})\overline{\mathcal{M}}_{7}\,,\\
  \mathcal{M}_{10}' &:=&  \overline{\mathcal{M}}_{10} + \frac{1}{3M_{\pi\Sigma}}(4E_{\Sigma}^{\ast}-m_{\Sigma})\overline{\mathcal{M}}_{11}\,,\qquad \mathcal{M}_{10}'' :=  \overline{\mathcal{M}}_{10} + \frac{1}{3M_{\pi\Sigma}}(4E_{\Sigma}^{\ast}+m_{\Sigma})\overline{\mathcal{M}}_{11}\,,\\
  \mathcal{M}_{14}' &:=&  \overline{\mathcal{M}}_{14} + \frac{1}{3M_{\pi\Sigma}}(4E_{\Sigma}^{\ast}-m_{\Sigma})\overline{\mathcal{M}}_{15}\,,\qquad \mathcal{M}_{14}'' :=  \overline{\mathcal{M}}_{14} + \frac{1}{3M_{\pi\Sigma}}(4E_{\Sigma}^{\ast}+m_{\Sigma})\overline{\mathcal{M}}_{15}\,.
\end{eqnarray*}
See Eq.~(\ref{eq:MmuDecomp}) for the definition of the $\mathcal{M}_{i}$, and Eq.~(\ref{eq:MbarApprox}) for the definition of the $\overline{\mathcal{M}}_{i}$. We point out that the structure functions $\mathcal{M}_{4,8,12,16}$ are always eliminated via the gauge-invariance constraints (\ref{eq:gaugeinv}).
\newpage
In the simple case of structure functions $\mathcal{M}_{i}$ independent of $\theta_{\Sigma}^{\ast},\,\phi_{\Sigma}^{\ast}$, vanishing for $i=3,7,11,15$, we have $\mathcal{M}_{i}'=\mathcal{M}_{i}''=\mathcal{M}_{i}$\,. In fact, if we wish, we can easily construct a gauge-invariant amplitude $\mathcal{M}^{\mu}$ of such a simplified form, which yields a set of prescribed $\mathcal{C}_{\ell\pm}^{i}(s,M_{\pi\Sigma}^2,t_{K})$,
\begin{eqnarray*}
  \mathcal{M}_{1}^{\mathcal{C}} &=& \frac{1}{4\sqrt{s}M_{\pi\Sigma}}\biggl((\sqrt{s}+M_{\pi\Sigma})\left((\sqrt{s}-m_{N})\mathcal{C}_{0+}^{1}+(\sqrt{s}+m_{N})\mathcal{C}_{1-}^{1}\right) \\ &\quad& \qquad\quad \,\,\,-\,\,\,(\sqrt{s}-M_{\pi\Sigma})\left((\sqrt{s}+m_{N})\mathcal{C}_{0+}^{2}+(\sqrt{s}-m_{N})\mathcal{C}_{1-}^{2}\right)\biggr)\,,\\
  \mathcal{M}_{2}^{\mathcal{C}} &=& \frac{1}{2\sqrt{s}M_{\pi\Sigma}}\biggl((\sqrt{s}+M_{\pi\Sigma})\left(\mathcal{C}_{0+}^{3}-\mathcal{C}_{1-}^{3}\right) + (\sqrt{s}-M_{\pi\Sigma})\left(\mathcal{C}_{0+}^{4}-\mathcal{C}_{1-}^{4}\right)\biggr)\,,\quad \mathcal{M}_{3}^{\mathcal{C}}=0\,,\\
  \mathcal{M}_{5}^{\mathcal{C}} &=& \frac{1}{4\sqrt{s}M_{\pi\Sigma}}\biggl((\sqrt{s}+M_{\pi\Sigma})\left(\mathcal{C}_{0+}^{1}-\mathcal{C}_{1-}^{1}\right) + (\sqrt{s}-M_{\pi\Sigma})\left(\mathcal{C}_{0+}^{2}-\mathcal{C}_{1-}^{2}\right)\biggr)\,,\\
  \mathcal{M}_{6}^{\mathcal{C}} &=& \frac{1}{2\sqrt{s}M_{\pi\Sigma}}\biggl(\frac{\sqrt{s}+M_{\pi\Sigma}}{\sqrt{s}+m_{N}}\left(\mathcal{C}_{0+}^{3}-\mathcal{C}_{0+}^{1}\right) + \frac{\sqrt{s}-M_{\pi\Sigma}}{\sqrt{s}-m_{N}}\left(\mathcal{C}_{0+}^{2}-\mathcal{C}_{0+}^{4}\right) \\ &\quad& \qquad\quad \,\,\,+\,\,\, \frac{\sqrt{s}+M_{\pi\Sigma}}{\sqrt{s}-m_{N}}\left(\mathcal{C}_{1-}^{3}-\mathcal{C}_{1-}^{1}\right) + \frac{\sqrt{s}-M_{\pi\Sigma}}{\sqrt{s}+m_{N}}\left(\mathcal{C}_{1-}^{2}-\mathcal{C}_{1-}^{4}\right)\biggr)\,,\quad \mathcal{M}_{7}^{\mathcal{C}}=0\,,\\
  \mathcal{M}_{9}^{\mathcal{C}} &=& -\frac{1}{4\sqrt{s}M_{\pi\Sigma}}\biggl((\sqrt{s}-m_{N})\left(\mathcal{C}_{0+}^{1}-\mathcal{C}_{1-}^{2}\right) + (\sqrt{s}+m_{N})\left(\mathcal{C}_{0+}^{2}-\mathcal{C}_{1-}^{1}\right)\biggr)\,,\\
  \mathcal{M}_{10}^{\mathcal{C}} &=& \frac{1}{2\sqrt{s}M_{\pi\Sigma}}\left(\mathcal{C}_{0+}^{4}-\mathcal{C}_{0+}^{3}+\mathcal{C}_{1-}^{4}-\mathcal{C}_{1-}^{3}\right)\,,\quad \mathcal{M}_{11}^{\mathcal{C}}=0\,,\\
  \mathcal{M}_{13}^{\mathcal{C}} &=& \frac{1}{4\sqrt{s}M_{\pi\Sigma}}\left(\mathcal{C}_{0+}^{2}-\mathcal{C}_{0+}^{1} + \mathcal{C}_{1-}^{2}-\mathcal{C}_{1-}^{1}\right)\,,\\
  \mathcal{M}_{14}^{\mathcal{C}} &=& \frac{1}{2\sqrt{s}M_{\pi\Sigma}}\left(\frac{\mathcal{C}_{0+}^{1}-\mathcal{C}_{0+}^{3} - \mathcal{C}_{1-}^{2} + \mathcal{C}_{1-}^{4}}{\sqrt{s}+m_{N}} - \frac{\mathcal{C}_{1-}^{1}-\mathcal{C}_{1-}^{3} - \mathcal{C}_{0+}^{2} + \mathcal{C}_{0+}^{4}}{\sqrt{s}-m_{N}}\right)\quad \mathcal{M}_{15}^{\mathcal{C}}=0\,,
\end{eqnarray*}
with $\mathcal{M}_{4,8,12,16}^{\mathcal{C}}$ accordingly fixed by the constraints (\ref{eq:gaugeinv})\,. - Finally, we define
\begin{eqnarray*}
  \mathcal{A}_{0+}^{1} &:=& \sqrt{E_{\Sigma}^{\ast}+m_{\Sigma}}\sqrt{E_{\pi\Sigma}+M_{\pi\Sigma}}\,\left(\mathcal{C}_{0+}^{1}\right)\sqrt{E_{N}-m_{N}}\,/\sqrt{2M_{\pi\Sigma}}\,,\\
  \mathcal{A}_{0+}^{2} &:=& \sqrt{E_{\Sigma}^{\ast}+m_{\Sigma}}\sqrt{E_{\pi\Sigma}-M_{\pi\Sigma}}\,\left(\mathcal{C}_{0+}^{2}\right)\sqrt{E_{N}+m_{N}}\,/\sqrt{2M_{\pi\Sigma}}\,,\\
  \mathcal{A}_{0+}^{3} &:=& \sqrt{E_{\Sigma}^{\ast}+m_{\Sigma}}\sqrt{E_{\pi\Sigma}+M_{\pi\Sigma}}\,\left(\mathcal{C}_{0+}^{3}\right)\sqrt{E_{N}-m_{N}}\,/\sqrt{2M_{\pi\Sigma}}\,,\\
  \mathcal{A}_{0+}^{4} &:=& \sqrt{E_{\Sigma}^{\ast}+m_{\Sigma}}\sqrt{E_{\pi\Sigma}-M_{\pi\Sigma}}\,\left(\mathcal{C}_{0+}^{4}\right)\sqrt{E_{N}+m_{N}}\,/\sqrt{2M_{\pi\Sigma}}\,,\\
  \mathcal{A}_{1-}^{1} &:=& \sqrt{E_{\Sigma}^{\ast}-m_{\Sigma}}\sqrt{E_{\pi\Sigma}+M_{\pi\Sigma}}\,\left(\mathcal{C}_{1-}^{1}\right)\sqrt{E_{N}+m_{N}}\,/\sqrt{2M_{\pi\Sigma}}\,,\\
  \mathcal{A}_{1-}^{2} &:=& \sqrt{E_{\Sigma}^{\ast}-m_{\Sigma}}\sqrt{E_{\pi\Sigma}-M_{\pi\Sigma}}\,\left(\mathcal{C}_{1-}^{2}\right)\sqrt{E_{N}-m_{N}}\,/\sqrt{2M_{\pi\Sigma}}\,,\\
  \mathcal{A}_{1-}^{3} &:=& \sqrt{E_{\Sigma}^{\ast}-m_{\Sigma}}\sqrt{E_{\pi\Sigma}+M_{\pi\Sigma}}\,\left(\mathcal{C}_{1-}^{3}\right)\sqrt{E_{N}+m_{N}}\,/\sqrt{2M_{\pi\Sigma}}\,,\\
  \mathcal{A}_{1-}^{4} &:=& \sqrt{E_{\Sigma}^{\ast}-m_{\Sigma}}\sqrt{E_{\pi\Sigma}-M_{\pi\Sigma}}\,\left(\mathcal{C}_{1-}^{4}\right)\sqrt{E_{N}-m_{N}}\,/\sqrt{2M_{\pi\Sigma}}\,,
\end{eqnarray*}
where $E_{\pi\Sigma}$ is the c.m. energy of the $\pi\Sigma$ pair, $E_{\pi\Sigma}=\sqrt{s}-E_{K}=(s+M_{\pi\Sigma}^2-M_{K}^2)/(2\sqrt{s})$. We also note that $|\vec{q}_{K}|=\sqrt{E_{\pi\Sigma}^2-M_{\pi\Sigma}^2}\,$ and $E_{N}=(s+m_{N}^2)/(2\sqrt{s})$\,.

\end{appendix}

\end{document}